\documentclass[twocolumn]{aastex62}
\usepackage{amsmath}
\usepackage{url}
\usepackage[percent]{overpic}

\newcommand{\spitzer}{\mbox {\em Spitzer}}
\newcommand{\chandra}{\mbox {\em Chandra}}

\newcommand{\um}{$\mu$m}
\newcommand{\hii}{\mbox{\ion{H}{2}~}}

\newcommand{\degree}{^{\circ}}
\newcommand{\Msun}{$M_{\sun}$}

\turnoffeditone
\turnoffedittwo

\accepted{for publication in {\em ApJ}, 2 June 2019}
\shorttitle{Duration of Star Formation in the Carina Nebula}
\shortauthors{Povich et al.}

\begin{document}

\title{The Duration of Star Formation in Galactic Giant Molecular Clouds.\\ I. The Great Nebula in Carina
}

\correspondingauthor{Matthew S. Povich}
\email{mspovich@cpp.edu}

\author{Matthew S. Povich}
\affiliation{Department of Physics and Astronomy, California State Polytechnic University, 3801 West Temple Ave, Pomona, CA 91768, USA}
\affiliation{Visitor in Astronomy, California Institute of Technology, Pasadena, CA 91125, USA}

\author{Jessica T. Maldonado}
\affiliation{Department of Physics and Astronomy, California State Polytechnic University, 3801 West Temple Ave, Pomona, CA 91768, USA}
\affiliation{Department of Physics and Astronomy, Michigan State University, East Lansing, MI 48824, USA}

\author{Evan Haze Nu\~{n}ez}
\affiliation{Department of Physics and Astronomy, California State Polytechnic University, 3801 West Temple Ave, Pomona, CA 91768, USA}

\author{Thomas P. Robitaille}
\affiliation{Freelance Consultant, Headingley Enterprise and Arts Centre, Bennett Road Headingley, Leeds LS6 3HN, United Kingdom}

\begin{abstract}
  We present a novel infrared spectral energy distribution (SED) modeling methodology that uses likelihood-based weighting of the model fitting results to construct probabilistic H--R diagrams (pHRD) for X-ray identified, intermediate-mass (2--8~\Msun), pre-main sequence young stellar populations. This methodology is designed specifically for application to young stellar populations suffering strong, differential extinction ($\Delta A_V > 10$~mag), typical of Galactic massive star-forming regions. We pilot this technique in the Carina Nebula Complex (CNC) by modeling the 1--8~\um\ SEDs of 2269 likely stellar members that \deleted{lack}\edit1{exhibit no excess emission from circumstellar dust disks at} 4.5~\um\ \deleted{excess}\edit1{or shorter wavelengths}. A subset of ${\sim}100$ intermediate-mass stars in the lightly-obscured Trumpler 14 and 16 clusters have available spectroscopic $T_{\rm eff}$, measured from the Gaia-ESO survey. We correctly identify the stellar temperature in \edit2{70}\% of cases, and the aggregate pHRD for all sources returns the same peak in the stellar age distribution as obtained using the spectroscopic $T_{\rm eff}$.  
The SED model parameter distributions of stellar mass and evolutionary age reveal significant variation in the duration of star formation among four large-scale stellar overdensities within the CNC \edit1{and} a large distributed stellar population. Star formation began ${\sim}10$~Myr ago and continues to the present day, with the star formation rate peaking ${\la}3$~Myr ago when the massive Trumpler 14 and 16 clusters formed. We make public the set of 100,000 SED models generated from standard pre-main sequence evolutionary tracks and our custom software package for generating pHRDs and mass-age distributions from the SED fitting results.
\end{abstract}


\section{Introduction}

The timescale over which giant molecular clouds (GMCs) collapse and produce new stars places fundamental constraints on theories of star formation \citep{MO07,K+12,Burkhart18} %[CITATIONS, dynamical vs. turbulent timescales],
calibrations of star formation rates \citep[SFRs;][]{CP11,KE12}, and the  dynamical evolution of \hii\ regions \citep{W+77,KM92,A+11,Z-A+19}.  Massive GMCs produce the O- and early B-type (OB) stars that dominate feedback on the galactic scale, hence measuring the duration of star formation in these structures calibrates the tachometers for the engines driving galaxy evolution \citep{Illustris,FIRE,FIRE-2}.

Our knowledge of constituent stellar populations in the most massive Galactic GMCs has been limited by the combination of relatively large heliocentric distances (${\ga}2$~kpc), relatively large angular extent (typically tens of arcmin to ${>}1\degree$ in diameter) high differential extinction ($\Delta A_V> 10$~mag), and overwhelming contamination by unassociated field stars in the Galactic plane \citep{CCCPOBc,MYStIXOBc}.
Large archival datasets from the {\it Chandra X-ray Observatory} and {\it Spitzer Space Telescope} have finally flung open an observational window to identify and resolve the young stellar populations associated with the most massive Galactic GMCs \citep{GLIMPSE,CCCP,MYStIX,MIRES,CygOB2,MOXC,MOXC2}. %[CITE MYStIX, MOXC, Cyg-X papers].

To constrain the duration of star formation, inter-mediate-mass (2--8~\Msun) stars are potentially the most valuable, yet still under-utilized, segment of the stellar mass distribution. OB stars are relatively rare and reach the zero-age main sequence (ZAMS) while still embedded in their dusty, natal cores \citep{ZY07}. The main-sequence turnoff for OB stars gives the age of individual massive clusters \citep[e.g.,][]{F+99,FCW05}, but cannot probe timescales shorter than the ${\sim}3$--20~Myr MS lifetimes of the most massive stars formed in a given cluster and breaks down if massive and low-mass stars are not coeval \citep{MH98}. Isochronal ages for low-mass, pre--main-sequence (pre-MS) stars are frequently employed, but these stars evolve relatively slowly along the fully-convective Hayashi tracks, requiring very precise placement on the Hertzsprung-Russell (HR) diagram to establish  precise ages. Distant GMCs produce additional challenges, as low-mass T Tauri stars are faint and difficult to detect, and their photospheres are frequently veiled by circumstellar dust in a disk and/or contaminated by accretion signatures such as strong emission lines \citep{S14}. By contrast, intermediate-mass, pre-MS stars (IMPS\footnote{This class of stars, the cooler progenitors of Herbig Ae/Be stars, have also been called intermediate-mass T Tauri stars \citep[IMTTS;][]{C+04} and G-type T Tauri stars \citep[GTTS][]{HS99}. We introduce the new acronym IMPS because it is more physically descriptive of cool, partially- or fully-convective stars\deleted{, including intermediate-mass protostars} \citep{PS91}. We hope to avoid further overworking the T Tauri moniker with additional modifiers, considering that two members of the eponymous triple system, T Tauri Sa and T Tauri N, each have masses of ${\sim}2$~\Msun \citep{K+16_TTau}, so they are themselves IMPS.}) evolve more rapidly than low-mass T Tauri stars. Some fraction of IMPS shed their circumstellar disks in 1--2~Myr \citep{HerbigAeBe,H07,P16} to reveal cool, convective photospheres before transitioning to the blue ZAMS in ${<}10$~Myr \citep{BM96,SDF00,H+19}. These stars have active dynamos powering magneto-coronal X-ray emission \citep{G+12,G16}, hence they are identifiable as {\it Chandra} point-sources even when the lack of an inner dust disk means there is no measurable mid-infrared (MIR) excess emission above the stellar photosphere. Although they may still possess debris disks, we will henceforth use the shorthand ``diskless'' to refer to X-ray selected young stars with MIR colors consistent with bare photospheres (for $\lambda \le 4.5~\mu$m).

%Be sure to mention here why we are focusing on this particular sample and acknowledge that our adoption of the term ``diskless'' is merely a convenient shorthand for lack of INNER dust disks, as these objects could certainly have transitional/debris disks that are undetectable against the bright nebular MIR background emission.

The wide-field, homogeneous X-ray and infrared datasets produced for the {\it Chandra} Carina Complex Project \citep[CCCP;][]{CCCP} provide an ideal test case for our new methodology. The Carina Nebula Complex (CNC) contains a
very large population of ${\ga}5\times 10^4$ stars arranged hierarchically among three major, massive clusters (Trumpler 14, 15, and 16), two dozen smaller clusters (including Collinder 228 and 232, Bochum 10 and 11, and the Treasure Chest), and as many as half of the young stars comprise a distributed population \citep{CCCP_xclust,K14_clust}. This spatial complexity reflects a complicated star-forming history over at least 10~Myr \citep{DG-E01}. The ages of individual subclusters within the CNC span at least several Myr \citep{FFM80_Tr15,Carraro02,Tapia+03,Smith+TC05,A+07_Tr14,Hur+12,AgeJX} and numerous regions of currently-active, possibly feedback-driven star formation permeate the remaining molecular cloud material \citep{SP10,CCCPYSOs,CarinaHerschel}. The CNC therefore boasts one of the largest, relatively nearby (${<}3$~kpc from the Sun) populations of IMPS, Herbig Ae/Be stars, and intermediate-mass ZAMS stars formed from a single GMC. Our methodology must succeed in reproducing the full range of ages represented by this population while handling the significant differential extinction (negligible $A_V$ in certain places, $A_V>15$~mag in others, with a reddening law that is known to vary with location and with increasing extinction; see \citealt{CCCPOBc} and references therein).

%AGE REFERENCES (just list them, discussed later)
%The idea that star formation in the CNC has been underway for ${\sim}10$~Myr is not new, having been proposed by . %Limited UBV photometric study of Tr 16, Tr 14 and one nearby, background field.

%\citet{FFM80_Tr15} found Tr 15 has $d=2.6$~kpc, age $6\pm 3$~Myr, from optical spectroscopy and $UBVRI$ photometry -- WOW! Spot on!

%MORE AGE REFERENCES FROM PREIBISCH+ 2011:
%Dias+ 2002 --- 8 Myr age for Tr 15, ~3 Myr for Tr 14 high-mass stars %These authors continue to revise their online catalog and have apparently made it WORSE by ingesting newer studies of questionable methods!
%\citep{Tapia+03} --- 3-40 Myr for Tr 15 (SO IT'S OLD), <1-5 Myr for Tr 14/16
%\citep{A+07_Tr14} --- 0-5 Myr for Tr 14, younger at center

%\citet{Carraro02} 6 Myr upper limit to Tr 15 from $UBVRI$ photometry.

\section{Data, Models, and Source Sample Construction}

\subsection{Source Datasets} \label{sec:data}

This pilot study uses previously-published CCCP datasets, specifically a large subset of the 4,664 \chandra\ X-ray point sources that were positionally matched to mid-infrared (MIR) sources from the {\em Spitzer Space Telescope} Vela-Carina Survey Point Source Archive \citep{CCCPCat,CCCPYSOs}.\footnote{Vela-Carina survey mosaics and point-source lists were produced using the GLIMPSE pipeline; for details of the processing and data products, go to \url{http://irsa.ipac.caltech.edu/data/SPITZER/GLIMPSE/doc/glimpse1_dataprod_v2.0.pdf}.}
The Archive provides broadband photometry data at 3.6, 4.5, 5.8, and 8.0 \um, plus $JHK_S$ near-IR (NIR) photometry from the {\em 2MASS} Point Source Catalog \citep{2MASS}, which is well-matched to both the 2\arcsec\ resolution of the \spitzer/IRAC detector \citep{IRAC} and the sensitivity limits of the shallow, \spitzer/GLIMPSE--style Legacy surveys ($[3.6] \la 15.5$ mag, \citealp{C09}). 

In the process of constructing the Pan-Carina YSO Catalog of \spitzer/IRAC MIR-excess sources, \citet{CCCPYSOs} identified 3,444 IR counterparts to CCCP X-ray sources that were consistent with normally-reddened stellar photospheres, plus 213 sources with ``marginal'' excess emission in the IRAC [5.8] or [8.0] bands. These 3,657 sources form the basis of our source sample, as they were possible X-ray detected members of the Carina Nebula Complex but did not exhibit significant 4.5~\um\ excess emission above a stellar photosphere. %Several additional classification and cleaning steps were required to produce our final sample of IR-bright, ``diskless'' members.

\subsection{Refinement of CCCP Membership List Using Gaia DR2 Parallaxes}\label{sec:parallax}
\citet{CCCPClass} classified 75\% of X-ray point sources in CCCP as probable members of the Carina Nebula young stellar population \edit1{based on proximity to observed spatial overdensities, X-ray brighness and median energy, and visual/infrared magnitudes}. \citet{K19_Gaia} analyzed proper motion and parallax information from Gaia DR2 \citep{GaiaDR2} for 28 young Galactic clusters and associations with available lists of X-ray selected members, including CCCP, and found that contamination from unrelated field stars was generally ${<}15\%$. We perform our own cleaning of the CCCP point-source catalog to remove remaining contaminants using an analysis of the Gaia DR2 parallax distribution, similar to that of \citet{K19_Gaia}. We do not consider proper motion information because we do not wish to impose a kinematic membership criterion that might exclude high-velocity stars, for example those ejected from one of the massive clusters but still found within the wide CCCP field-of-view.

We first identified Gaia DR2 matches to CCCP X-ray sources with 2MASS counterparts \citep{CCCPCat}. The nearest Gaia DR2 source falling within 1.2\arcsec\ of the 2MASS source was declared matched. We then analyzed the parallaxes of 4053 Gaia DR2 matches with $G > 8$~mag and $\texttt{astrometric\_excess\_noise} < 1.0$~mas. The uncertainty on the parallax for each source was adjusted from the DR2 catalog value using the ``tentative external calibration'' of \citet{Gaia_astrom18}.
% [CITATION: Lindegren et al. 2018, A\&A, 616, A2].
We then computed the median of the parallax distribution, omitting 417 sources that were ${>}1\sigma$ outliers. Our median parallax $\varpi_0=0.40\pm 0.04$~mas gives a distance of $2.50^{+0.28}_{-0.23}$~kpc, in agreement with \citet{K19_Gaia} within the uncertainties, which are dominated by our shared assumption of a 0.04~mas systematic uncertainty in the parallax zeropoint. This moves the Carina nebula to a larger distance than the 2.3~kpc assumed in the original CCCP studies, however this distance value is still consistent with the lower end of the Gaia DR2 distance estimates. Finally, we flagged 294 CCCP sources with individual parallax measurements ${>}3\sigma$ above the median (based on their individual, adjusted parallax uncertainties) as foreground stars. Of these, 119 were previously classified as probable members of the Carina complex by \citet{CCCPClass}.
None of the CCCP sources could be analogously flagged as a background star.

Our parallax-based cleaning of probable foreground stars thus removed 7\% of all IR-bright CCCP sources.
Field stars that were undetected by Gaia or detected but with sufficiently high DR2 parallax uncertainties that they could not be confidently removed may persist as residual contaminating sources.

\subsection{The SED Models}

For this and future papers in this series we employ a set of $10^5$ ``naked'' pre-MS spectral energy distribution (SED) models that was first described by \citet[][hereafter P16]{P16}. These models are publicly-available.\footnote{\url{https://doi.org/10.5281/zenodo.2647586}}
The naked pre-MS model set is similar to the \texttt{s-s-i} model set from \citet[][hereafter R17]{R17}. Both of these model sets consist of spherical stellar photospheres only, with no circumstellar dust in a disk, envelope, or ambient medium. 
The naked pre-MS models use \citet{Kurucz} plane-parallel LTE stellar photospheres, the same as the R17 models for $T_{\rm eff}\ge 4,000$~K. 
%There is a key difference in how these models are defined and sampled.
All R17 model sets describe the central sources solely in terms of stellar radius $R_{\rm eff}$ and temperature $T_{\rm eff}$, with no reference to a particular set or evolutionary tracks and hence no assignment of stellar mass or age. By contrast, for the naked pre-MS models we sampled the stellar mass $M_{\star}$ and age $t_{\star}$ were sampled uniformly in logarithmic space \citep[][P16]{grid} and then converted to ($T_{\rm eff}$,$R_{\rm eff}$) space using pre-MS evolutionary tracks \citep{BM96,SDF00}. This adds a third, derived parameter variable, surface gravity, to the naked pre-MS models that is absent in the \texttt{s-s-i} models, but in practice broadband photometry only constrains the independent variables $R_{\rm eff}$ and $T_{\rm eff}$.

%%%%%%%%% FIGURE 1 %%%%%%%%%%%
\begin{figure}[tph]
%\epsscale{1.0}
\includegraphics[width=\columnwidth]{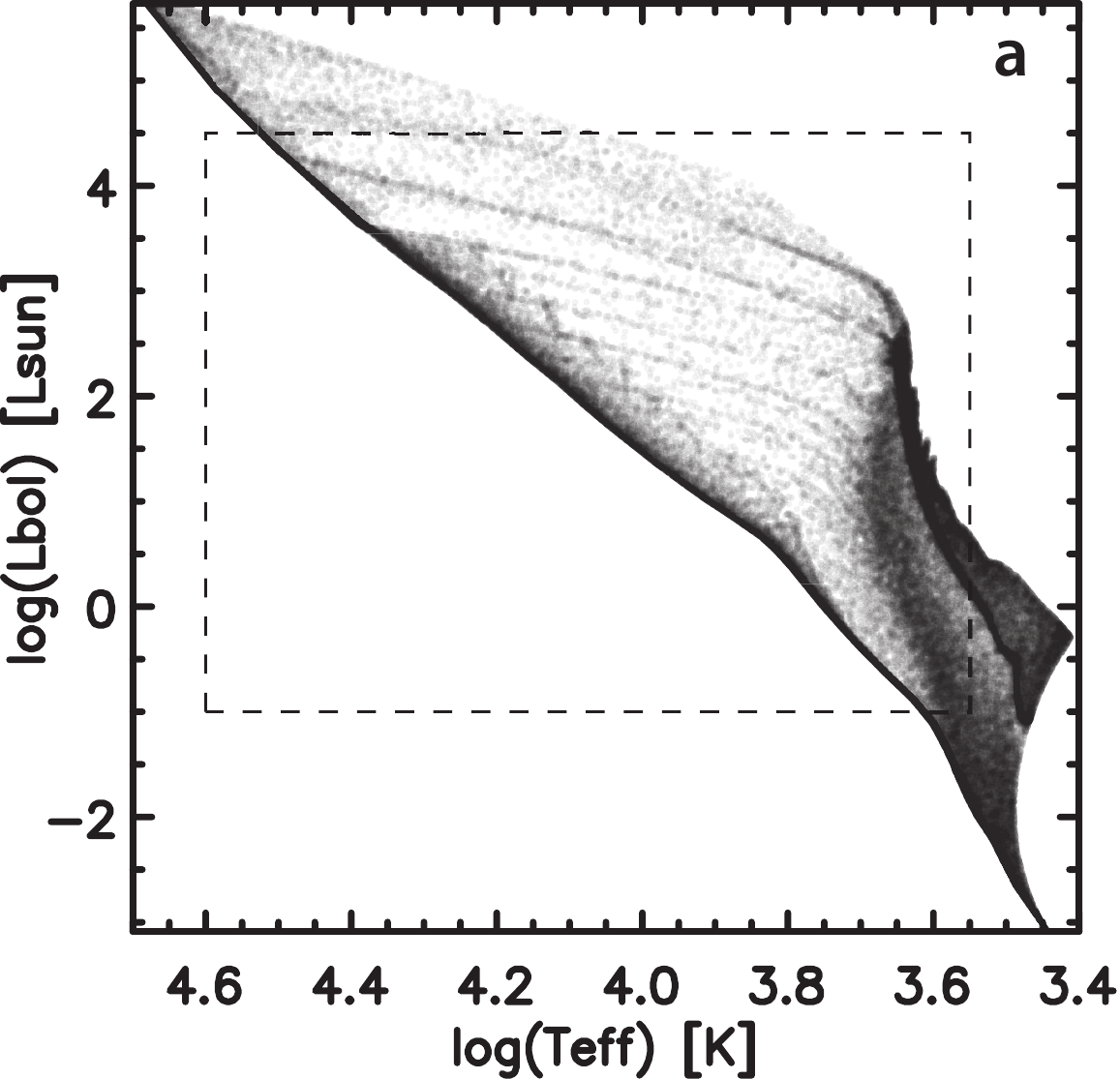}
\includegraphics[width=\columnwidth]{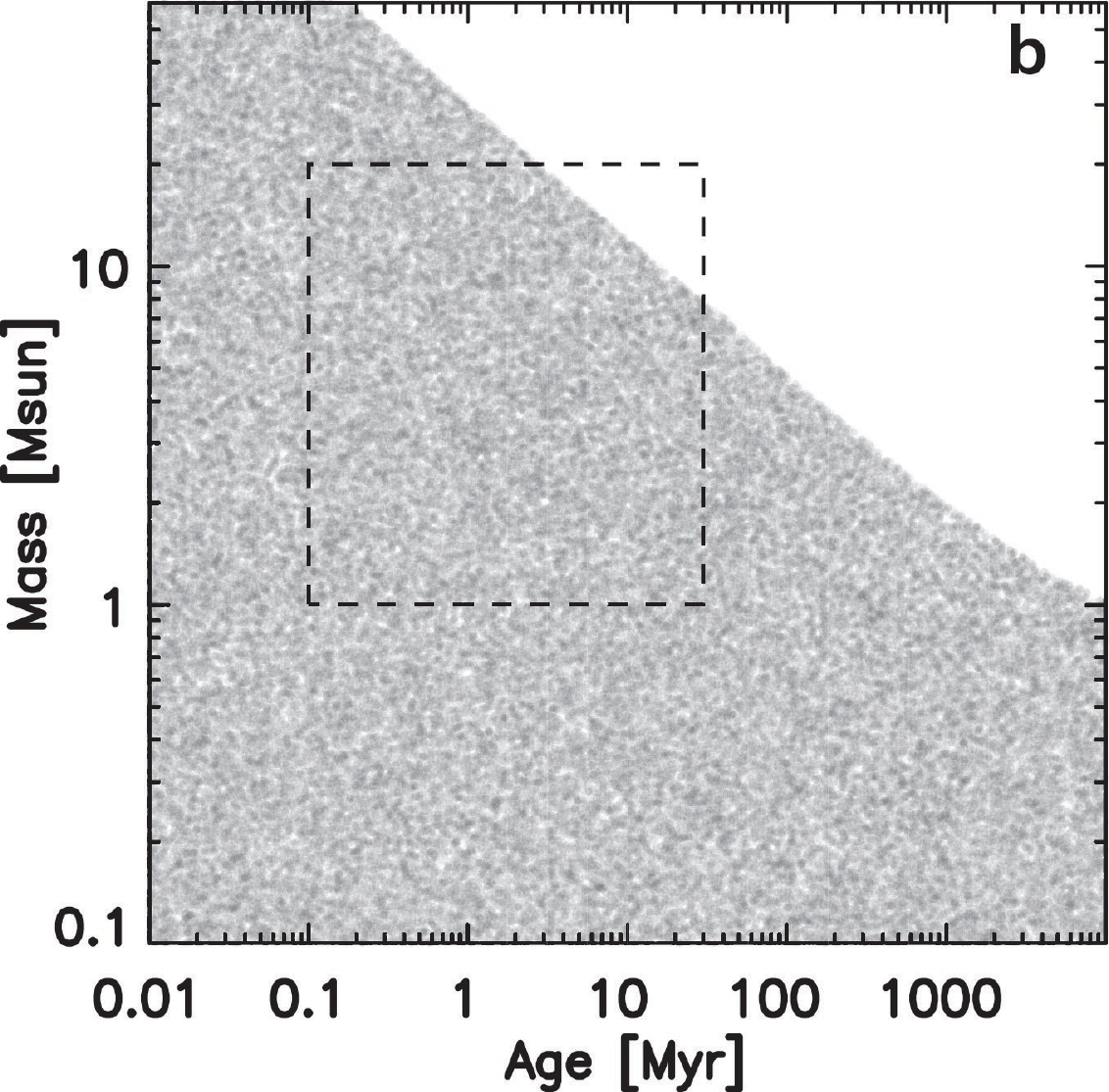}
\caption{Plots showing parameter spaces sampled by the $10^5$ naked stellar models. Each model is plotted as a single, grey dot, with transparency such that areas with high density of overlapping models turn darker grey to black. (a) Bolometric luminosity versus effective temperature, replicating a traditional H--R diagram. \edit1{The striping structure visible in the model density is an artifact of the interpolation between the discrete set of pre-PM evolutionary tracks.} (b) Stellar mass versus evolutionary age, the fundamental parameter plane used to construct the grid. Dashed boxes on both panels show the boundaries of the parameter ranges of interest for the diskless, IR-bright sources in this study.
\label{fig:params}}
\end{figure}

The regions of the traditional Hertzsprung-Russell diagram (HRD; $L_{\rm bol}$--$T_{\rm eff}$ space) and of $M_{\star}$--$t_{\star}$ space sampled by the naked pre-MS models are illustrated in Figure~\ref{fig:params}.
For our goal of constraining age distributions of large stellar populations, both this parameter sampling and the larger size of the naked pre-MS models offer key advantages over the R17 \texttt{s-s-i} models. The uniform sampling in $\log{t_{\star}}$ maps to the familiar, highly non-uniform distribution on the HRD characterized by the thin, densely-populated diagonal line of the ZAMS and the nearly-vertical pre-MS overdensity. The density of models is greatly reduced between the ZAMS and the pre-MS, a region that has been called the R-C gap \citep{M+07} because it falls between intermediate-mass main sequence stars with radiative envelopes and their pre-MS analogs that are still fully or partially convective. The region of the HRD to the lower-left of the ZAMS is unphysical and hence unpopulated (Figure~\ref{fig:params}a). These models do not include post--main-sequence evolution, creating an unpopulated region in the upper-right of the mass--age parameter space (Figure~\ref{fig:params}b). While much  effort continues to be devoted to the further development and refinement of pre-MS evolutionary models \citep[e.g.,][]{SP15,Choi+16_MIST,Dotter16_MIST,H+19}, all stellar populations synthesized from pre-MS evolutionary tracks share these same general, qualitative features. The naked pre-MS SED models therefore possess a built-in, physical ``prior'' probability distribution of where observed young stars are most likely to be placed on the HRD. This comes at the cost of assuming a particular set of pre-MS evolutionary tracks that may not be correct. The evolutionary models used in our SED models are two decades old, but we adopt them because (1) they have been widely used in the literature, particularly in studies of isochronal ages in the CNC and other Galactic massive star-forming regions to which we will directly compare the results from our new method; and (2) most recent innovations in pre-MS evolutionary models have focused on low-mass stars. We discuss how the choice of different evolutionary models could impact our results in Section~\ref{sec:evochoice}.

%%%%%%%%% FIGURE 2 %%%%%%%%%%%
\begin{figure}[tbp]
%\epsscale{1.0}
\includegraphics[width=\columnwidth]{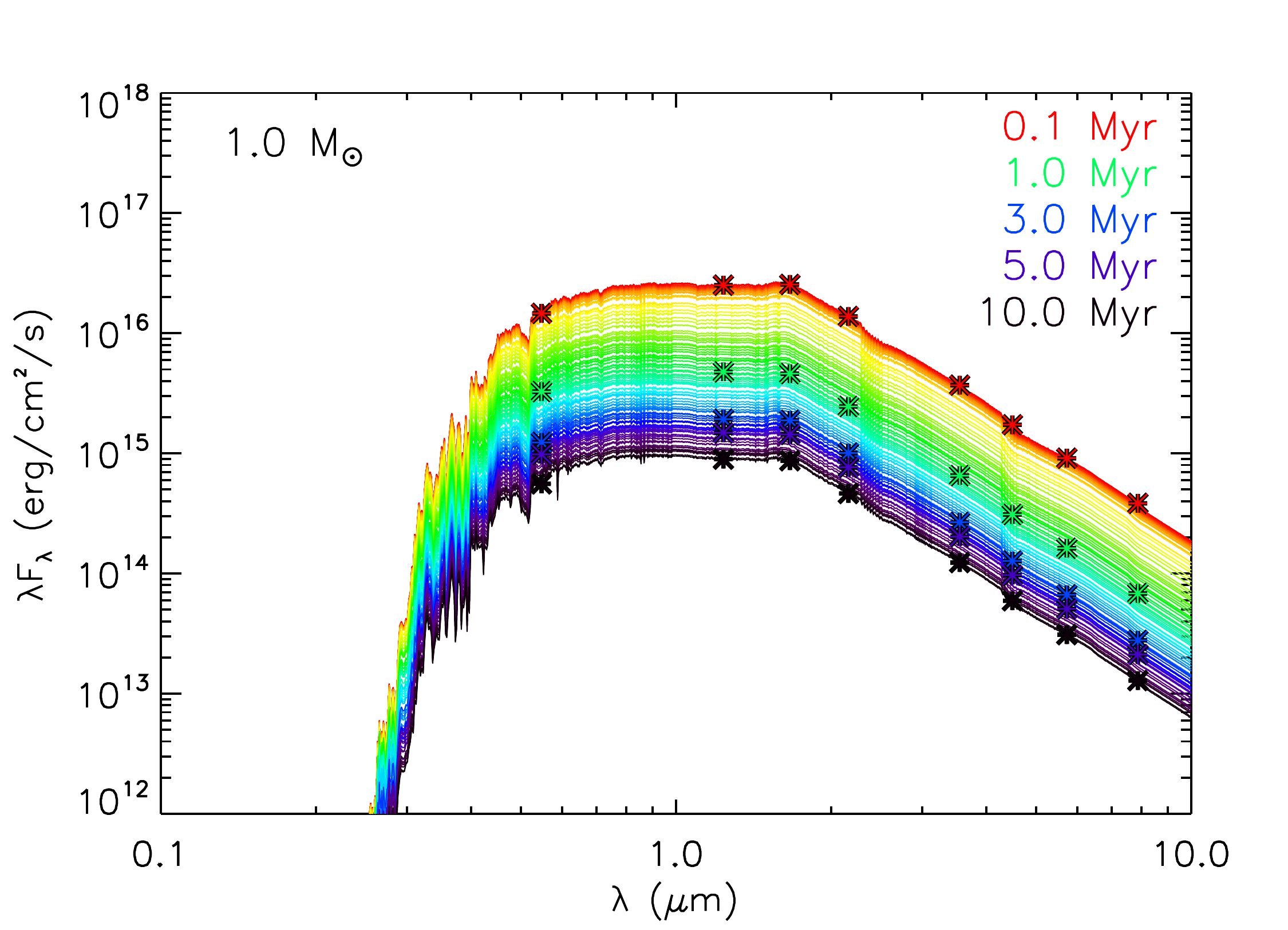}
\includegraphics[width=\columnwidth]{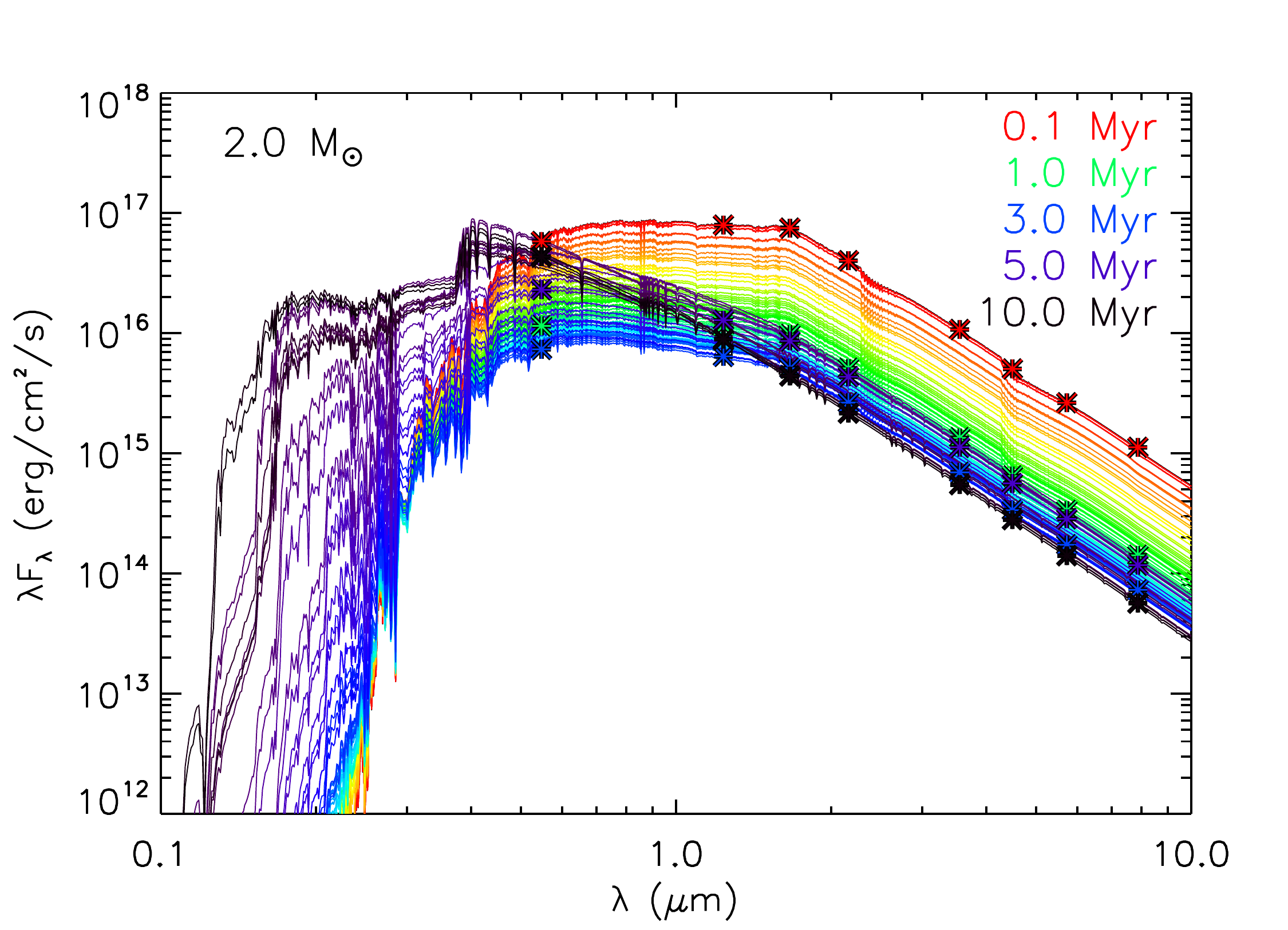}
\includegraphics[width=\columnwidth]{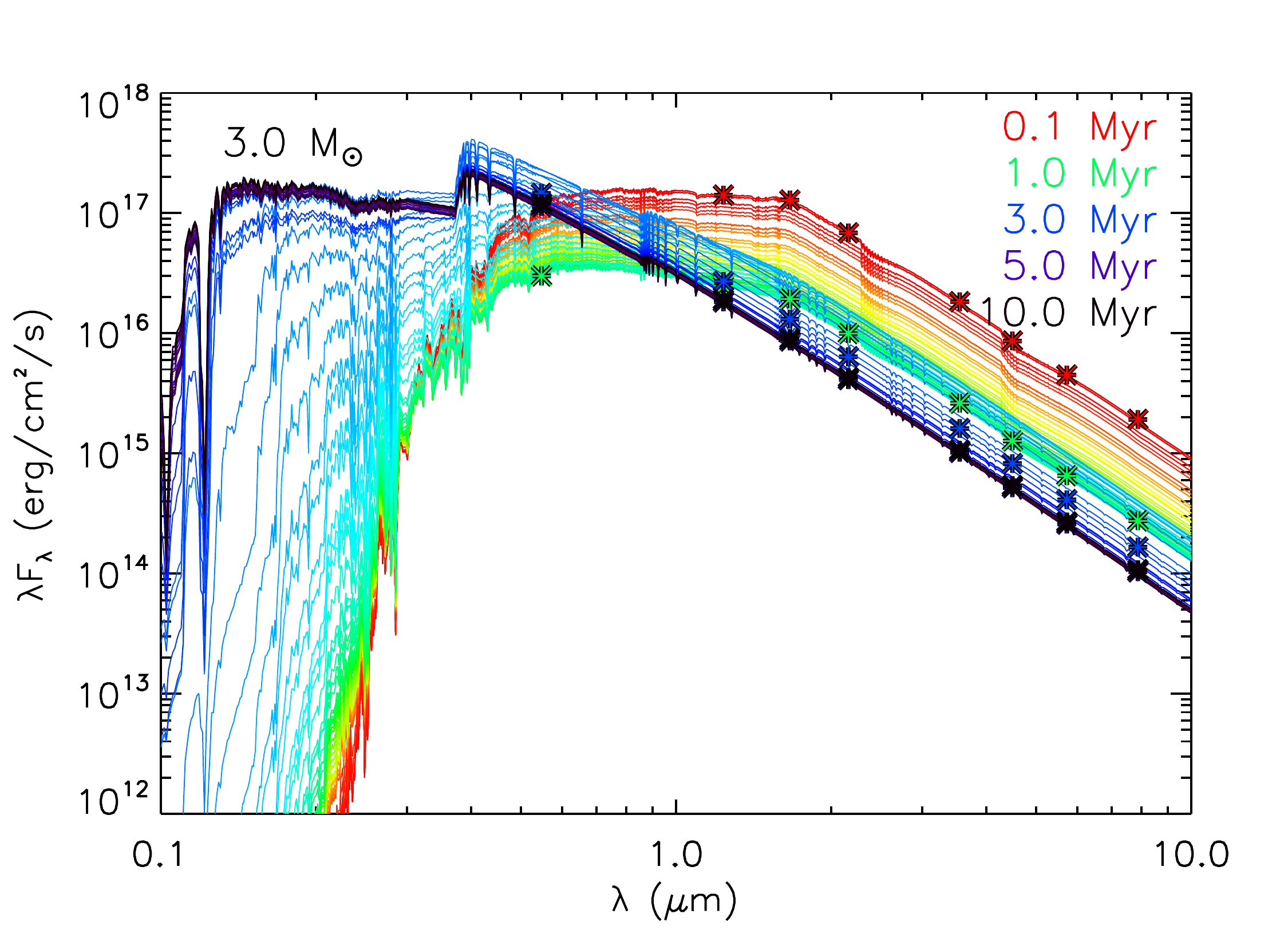}
\caption{Time-evolution of the SED models for the first 10~Myr along the 1, 2, and 3~\Msun\ \citet{SDF00} tracks. The 100 spectra plotted in each panel are \citet{Kurucz} atmospheres, with stellar temperatures and radii sampled at logarithmic age intervals and color coded from red (youngest) through dark violet to black (oldest). \edit1{Model fluxes convolved with the $VJHK_s$ and IRAC bandpasses are overplotted as color-coded asterisks for each  isochronal age indicated in the plot legends.}}
\label{fig:SEDs}
\end{figure}
Figure~\ref{fig:SEDs} illustrates the time-evolution of the model SEDs for a 1, 2, and 3~\Msun\ star over the first 10~Myr. The 1~\Msun\ model star shows no measurable change in SED shape over this timescale, typical of low-mass pre-MS stars that slowly descend the Hayashi tracks at near-constant $T_{\rm eff}$. The intermediate-mass model SEDs, by constrast, exhibit a rapid transition from cool, pre-MS stars to hot, A or B-type ZAMS stars (between 3 and 5~Myr for the 2~\Msun\ evolutionary track and between 1 and 3~Myr for 3~\Msun). This change in SED shape is measurable using broadband IR photometry. Specifically, $JHK_S$ photometry can distinguish between the cool pre-MS and hot ZAMS SEDs, constraining $T_{\rm eff}$, while  \spitzer/IRAC photometry at ${\ge}3.6$~\um\ places strong constraints on the Rayleigh-Jeans tail of the SED and hence $R_{\rm eff}$ (or $L_{\rm bol}$) for a specified $T_{\rm eff}$. 

\subsection{Fitting Models to the Data}\label{sec:fitting}

We fit the SED models to available broadband photometry data using the publicly-available Python port (R17)\footnote{\url{http://doi.org/10.5281/zenodo.235786}} of the \citet{fitter} SED fitting tool. All of these CCCP-matched IR sources had been previously fit with different SED models and found consistent with stellar photospheres by \citet{CCCPYSOs}, which imposed the requirement that all sources were detected in at least four of the seven combined 2MASS and IRAC photometry bands, including both the IRAC [3.6] and [4.5] bands. The majority of sources were detected in all five photometry bands from 1--4.5~\um. Because the SED declines toward longer IR wavelengths for these diskless sources and the Carina Nebula itself produces very bright nebular IR background, detection in the IRAC [5.8] and [8.0] bands was less frequent. For the ${\sim}10\%$ of sources with ``marginal'' excess emission detected at [5.8] or [8.0] these bands were not used for SED fitting.

The extinction curve of \citet{I05} was applied to the SED models as part of the fitting process, which introduced a third, independent parameter variable, $A_V$, to our fitting results. Using the standard $J-H$ versus $H-K_S$ color-color diagram we determined the maximum extinction to our IR-bright CCCP sources to be $A_V=15$~mag and imposed this as a hard upper limit for reddened SED models. We also constrained the scale parameter $R_{\rm eff}/d$ of the model SEDs by restricting the distance to $2.3~{\rm kpc}\le d \le 2.8$~kpc.

Following our well-established practice, we use the (unreduced) $\chi^2$ goodness-of-fit parameter normalized to the number of photometry data points used in the fit ($N_{\rm data}$) to identify sources that can be successfully modeled with our chosen model set \citep[e.g.][]{CCCPYSOs,MIRES,R17}. Because the naked SED models are new, we experimented with different values and found well-fit sources to be those for which the single best-fit model satisfied $\chi^2_0/N_{\rm data}\le 1$. We explain our rationale for this choice of cutoff value in Section~\ref{sec:grading} below.

\subsection{Likelihood-Based Weighting of SED Model Fits}\label{sec:weighting}
We define the set $i$ of well-fit SED models to each source using
\begin{equation}
  \frac{\chi^2_i-\chi^2_0}{N_{\rm data}}\le 1.
\end{equation}
This is a more strict cutoff than was used by P16, but still typically generates sets of hundreds or even thousands of well-fit model parameters for a given source.

Interstellar dust makes the photometric colors of reddened, hotter stars nearly indistinguishable from those of unreddened, cooler stars, an unfortunate property of nature that greatly complicates observational stellar astronomy. This manifests itself in our SED modeling as a strong degeneracy between $T_{\rm eff}$ and $A_V$. To ameliorate the impact of this degeneracy on our results, we leverage external information about each source in our sample to weight the relative likelihood of each model in the well-fit set.\footnote{The software routines and step-by-step procedures used to carry out our advanced analysis of the SED fitting results from this point onwards are publicly-available as an IDL library at \url{https://doi.org/10.5281/zenodo.3234101}. 
We plan to produce a \texttt{python} port in the future.} As described in Appendix B of P16, the relative probability of each individual model $i$ in the set of well-fit models is given by
\begin{displaymath}
  P_i(\tau_d,\tau_X) = P_nW_i(\chi^2)W_i(\tau_d)W_i(\tau_X)
\end{displaymath}
(their Equation 5), in which $P_n$ is a normalization constant ensuring that $\sum{P_i}=1$. We compute the first two of the following weighting terms using the functions defined by P16.
The likelihood of each model based on goodness-of-fit is
  $W_i(\chi^2)$ %= e^{-\chi^2_i/2}
(Equation 4 of P16).
The likelihood that a star of a particular mass and age is diskless
is $W_i(\tau_d)$, computed for each $(M_{\star}, t_{\star})$ parameter combination  using Equations 6 and 7 of P16. This weighting function disfavors very young, naked SED models for which we would expect the MIR SED to include emission from a circumstellar dust disk or protostellar envelope. The $W_i(\tau_d)$ weighting term is only appropriate for sources where we can be confident that the stellar photosphere is dominates the detected MIR emission at 4.5~\um.

The final weighting term, $W_i(\tau_X)$, is the likelihood that the star is X-ray detected based on the model $(M_{\star}, t_{\star})$ parameter combination. This term attempts to quantify our knowledge that bright, coronal X-ray emission is a strong indicator of youth. P16 cautioned that their weighting function was strictly appropriate only for low-mass, T Tauri stars for which coronal X-ray emission has been well-characterized.
However, \citet{G16} reported coronal X-ray emission for pre-MS stars of all masses in a sample that included stars up to 3~\Msun, the progenitors of main-sequence A-type stars. These authors also demonstrated that the X-ray luminosity decreases more rapidly with age in more massive pre-MS stars, following the time elapsed since the development of a radiative core in partially-convective stars. Using the $L_X(t)$ functional relationships for various mass ranges provided by \citet{G16}, we have revised the P16 weighting function based on X-ray detection as follows:
\begin{equation}\label{eq:xweight}
W_i(\tau_X) = 
\begin{cases}
  \left(\frac{t_{\star,i}}{\tau_X}\right)^{-\beta_X}, & t_{\star,i} > \tau_X \\
  1, & t_{\star,i} \le \tau_X.
\end{cases}
\end{equation}
We have introduced two new mass-dependent parameters, $\tau_X$ and $\beta_X$. The critical age at which a radiative core first develops is
\begin{equation}\label{agecritx}
  \frac{\tau_X(M_{\star})}{\rm yr} = 
  \begin{cases}
    10^6\left(\frac{1.494}{M_{\star}}\right)^{2.364}, & 1 < M_{\star}/M_{\odot} \le 15 \\
      2.583\times 10^6, & M_{\star}/M_{\odot} \le 1.
  \end{cases} 
\end{equation}
Once a pre-MS star evolves beyond the critical age, its X-ray emission decays as a power-law governed by
\begin{equation}\label{betax}
  \beta_X =
  \begin{cases}
    1.19, & 2 < M_{\star}/M_{\odot} \le 15 \\
    0.86, & 1.5 < M_{\star}/M_{\odot} \le 2 \\
    0.75, & M_{\star}/M_{\odot} \le 1.5. 
    \end{cases}
\end{equation}
For low-mass T Tauri stars, $\beta_X$ is identical to the value reported by \citet{PF05} and adopted by P16. For the IMPS, however, the much more rapid decay in X-ray emission provides significantly stronger $W_i(\tau_X)$ constraints on the SED model fitting results.

%%%%%%%%% FIGURE 3 %%%%%%%%%%%
\begin{figure*}[thp]
%%\epsscale{0.49}
\centering
\begin{overpic}[width=\linewidth]{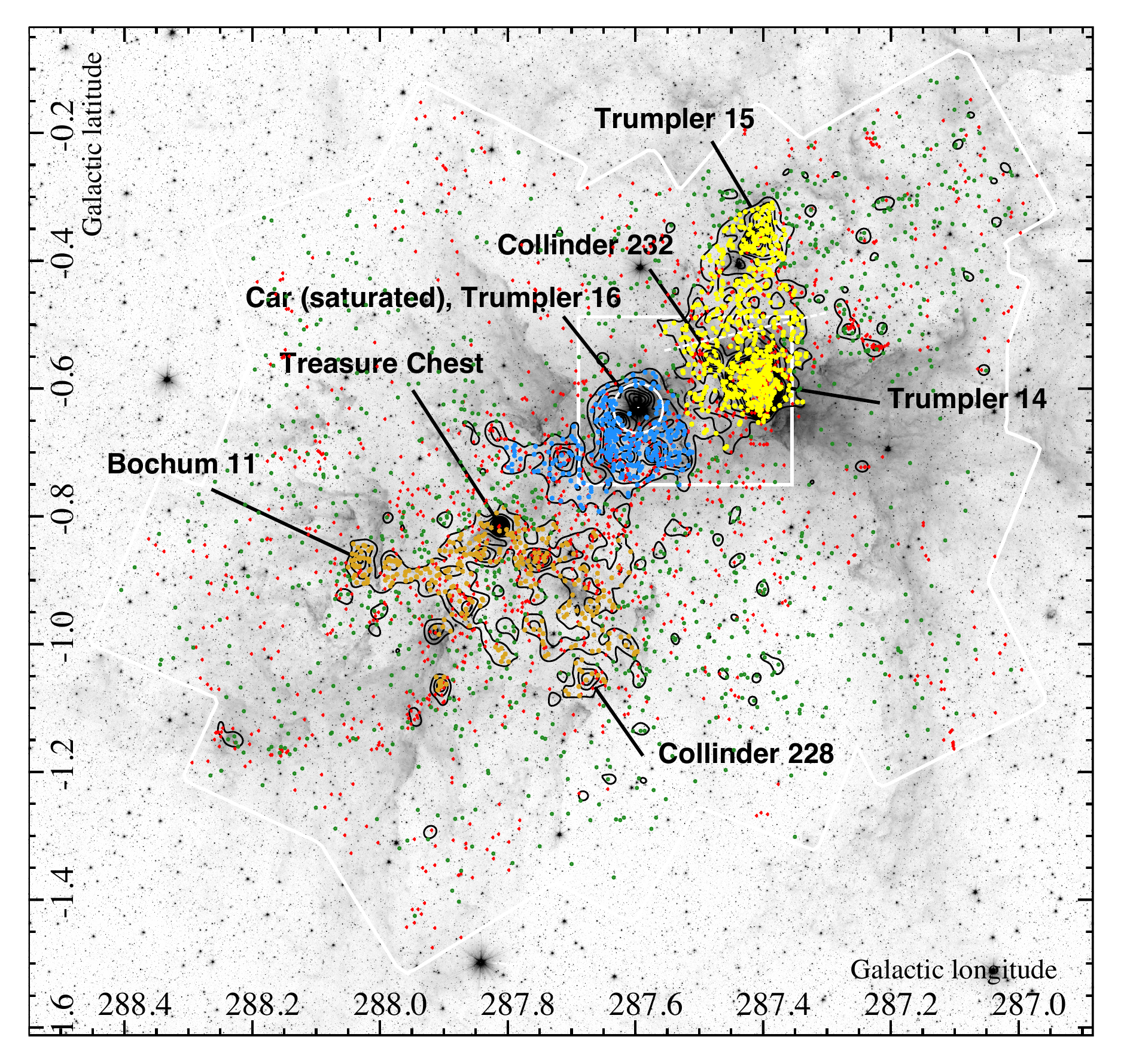}
  \put (20,67.35) {\large$\eta$}
\end{overpic}
\caption{\spitzer/IRAC 3.6~\um\ mosaic image (inverted logarithmic grayscale) with small circles overplotted to mark locations of the intermediate-mass stellar population classified via IR SED fitting. Black contours trace the density of X-ray-identified probable members and reveal three principal sub-regions with large-area spatial overdensities \citep{CCCP_xclust}. The 2269 diskless sources are color-coded according to their sub-region associations: {\em yellow} = Region A (including Tr 14 and 15 and Coll 232, we further separate these sources into regions A1 and A2 above and below dashed, white line); {\em blue} = Region B (including Tr 16); {\em orange} = Region C (the South Pillars, including Bo 11, Coll 228, and the Treasure Chest); and {\em green} = Region D (the distributed population). Disk-bearing YSOs from the Pan-Carina YSO Catalog \citep{CCCPYSOs} are overplotted as 1432 small red circles; these IR excess sources are not used in our analysis. IR point-source sensitivity is compromised within a ${\sim}1\arcmin$ radius of the strongly saturated source $\eta$~Car (white circle).  The white box encloses the zoomed-in region shown in Figure~\ref{fig:zoom} while the outer white outline marks the boundaries of the CCCP X-ray survey area \citep{CCCP}. 
\label{fig:overview}
} 
\end{figure*}

\subsection{The CCCP IR-Bright, Diskless Source Sample}

After selecting only probable members from \citet{CCCPClass}, cleaning out remaining foreground contaminants using Gaia DR2 parallaxes, and imposing our goodness-of-fit criterion, our final sample contains 2269 IR-bright, diskless stellar counterparts to CCCP X-ray point sources.
The spatial distribution of these probable intermediate-mass members of the CNC young stellar population is shown in Figure~\ref{fig:overview}. The spatial distribution is complex and hierarchical, exhibiting sub-clustering on multiple size scales \citep{K14_clust,Buckner+19}.
\citet{CCCP_xclust} divided the clustered CCCP source population into three principal spatial overdensities (regions A, B, and C, from northwest to southeast) and assigned the remaining sources to a distributed population (region D).
To study large-scale variations in the star formation history across the complex we assign each source in our sample to one of these regions, shown using different-colored symbols enclosed by the lowest density contours in Figure~\ref{fig:overview}. Region A contains the massive Tr 15 and 14 clusters,  which are known to have very different ages \citep[e.g.,][]{FFM80_Tr15,A+07_Tr14}, so we bisected this
into two sub-regions, A1 and A2 (separated by the dashed line in Figure~\ref{fig:overview}). This yielded ${\ga}300$ sources per sub-region (1050 sources in the distributed population), sufficiently large samples for robust statistical analysis of the SED fitting results in each.

\section{The Gaia-ESO Spectroscopic Comparison Sample}
%%%%%%%%%% FIGURE 4: Zoom-in on Tr 16 and 14 region with Comparison Sample Overlaid %%%%%%%%%%%
\begin{figure*}[thbp]
%%\epsscale{0.49}
\centering
\begin{overpic}[scale=0.8]{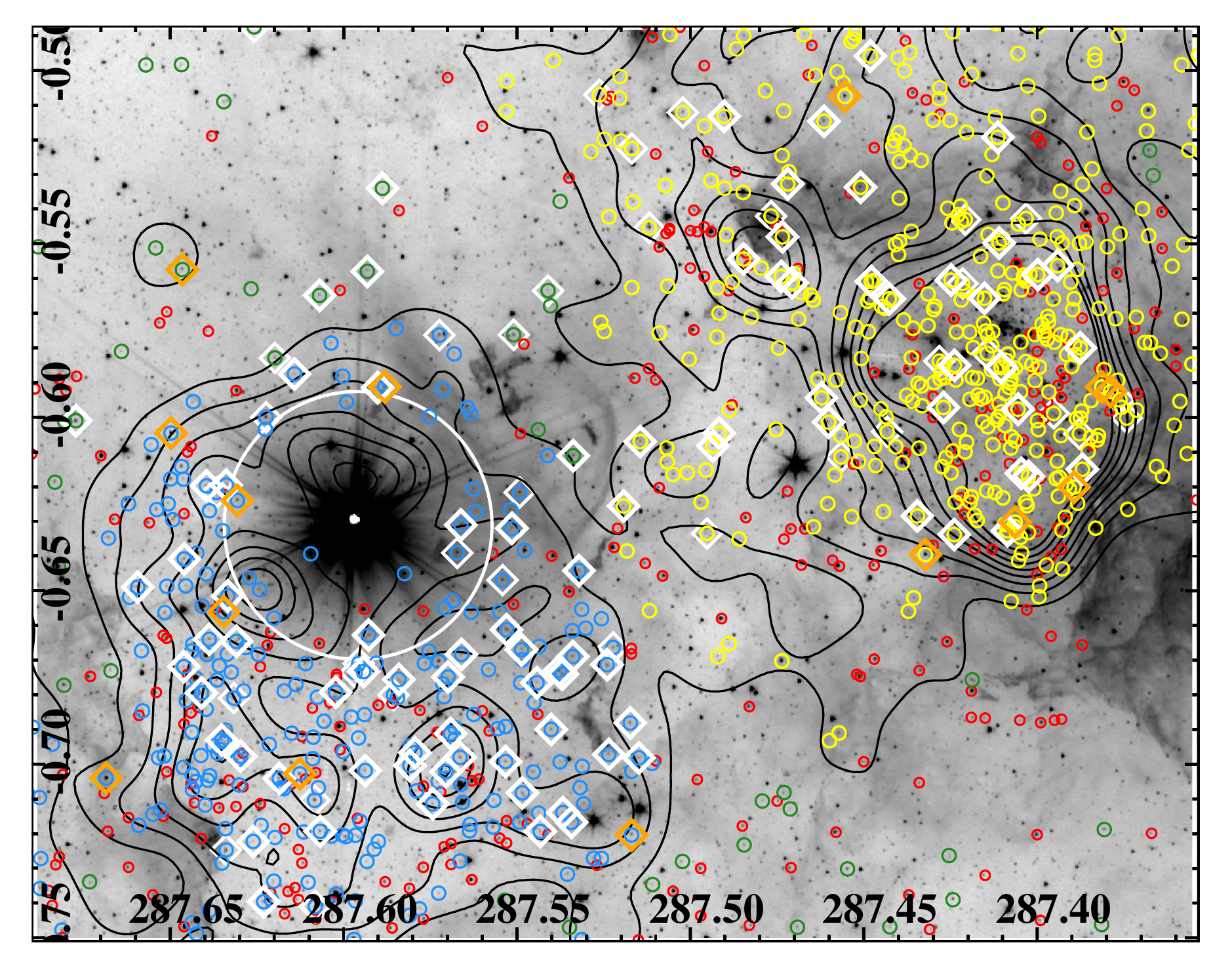}
  \put (40,-2) {\large {\bf Galactic longitude}}
  \put (-2,37) {\rotatebox[origin=c]{90}{\large {\bf Galactic latitude}}}
\end{overpic}
  \vspace{1 em}
\caption{Zoomed-in image of the central, lightly-obscured region of the Carina Nebula, containing the massive Tr 16 (left) and Tr 14 (right) clusters plus the smaller Collinder 232 cluster (immediate upper-left of Tr 14). Diamonds mark the Gaia-ESO counterparts from D17 matching our diskless population ({\em white} = late-type stars and {\em orange} = B-type stars). Other overlays are the same as in Figure~\ref{fig:overview}.
\label{fig:zoom}
} 
\end{figure*}

%%%%%% TABLE 1: Explanation of Individual Grades %%%%%%%%%%%%
\begin{table*}[htb]
  \begin{center}
  \caption{ \label{tab:grades}
    Summary of $T_{\rm eff}$ accuracy grades among the Gaia-ESO comparison sample}
%  \begin{tabularx}{\textwidth}{crrX}
  \begin{tabular}{crrl}
    Grade & \edit1{$N_{\rm IR}$} & \edit1{$N_{\rm V+IR}$} & Definition \\
    \hline\hline
    A & \edit1{23} & \edit1{22} & Tightly-constrained $T_{\rm eff}$ parameter \edit1{($\sigma_T/\overline{T_{\rm eff}}\le 0.05$); $\overline{T_{\rm eff}}$ within 10\% of $T_{\rm eff,S}$.} \\
    B & \edit1{28} & \edit1{40} & Well-constrained  $T_{\rm eff}$ parameter \edit1{($\sigma_T/\overline{T_{\rm eff}}\le 0.1$); $\overline{T_{\rm eff}}$ within 30\% of $T_{\rm eff,S}$.}  \\
    C & \edit1{32} & \edit1{19} & Poorly-constrained $T_{\rm eff}$ parameter \edit1{($\sigma_T/\overline{T_{\rm eff}}> 0.1$); $\overline{T_{\rm eff}}$ within 30\% of $T_{\rm eff,S}$.} \\
    D & \edit1{32} & \edit1{25} & Poorly-constrained $T_{\rm eff}$ parameter \edit1{($\sigma_T/\overline{T_{\rm eff}}> 0.1$); $\overline{T_{\rm eff}}$ deviates ${>}30\%$ from $T_{\rm eff,S}$.} \\
    F & \edit1{10} & \edit1{8} & \edit1{Well-constrained $T_{\rm eff}$ parameter  ($\sigma_T/\overline{T_{\rm eff}}\le 0.1$);  $\overline{T_{\rm eff}}$ deviates ${>}30\%$ from $T_{\rm eff,S}$.} \\
    \hline
    [Early] & 14 & \edit1{11} & Spectra and SED modeling agree on B-type star (AB grade equivalent). \\
    \hline
  \end{tabular}
%  \end{tabularx}
  \end{center}
  \edit1{\tablecomments{$\overline{T_{\rm eff}}$ and $\sigma_T$ are the $P_i(\tau_d,\tau_X)$--weighted mean and standard devation, respectively, of the $T_{\rm eff}$ parameter distribution for each source.}}
\end{table*}

\citet[][hereafter D17]{D17} analyzed a large set of spectra, obtained as part of the Gaia-ESO survey \citep{Gaia-ESO}, for 1085 stars in a relatively small, lightly-obscured field containing the massive Trumpler (Tr) 14 and 16 clusters (white box in Figure~\ref{fig:overview}). This stellar sample was selected using the optical photometric catalog of \citet{Hur+12}. D17 used a combination of CCCP X-ray detection, radial velocity, and Lithium equivalent width (Li EW) $> 150$~m\AA\ to identify 286 likely Carina members with spectral types A or later. D17 published the spectroscopically-determined \edit1{$T_{\rm eff,S}$} and uncertainties for all of these stars. Of these, 125 were also found within our IR-bright, diskless sample of CCCP sources (white diamonds in Figure~\ref{fig:zoom}). 

D17 also identified 110 early-type members for which no $T_{\rm eff,S}$ measurements were possible. Two of these had spectra consistent with obscured O-type stars (previously reported as candidate X-ray emitting O stars by \citealt{CCCPOBc}) while the rest were likely B-type stars. A large majority of the B-type stars were not detected as CCCP sources, presumably because they are X-ray quiet, and only 14 are found in our sample (orange diamonds in Figure~\ref{fig:zoom}).\footnote{Because D17 did not tabulate CCCP associations for the small minority of their early-type stars that had them, we instead cross-matched their list to our sources using the 2MASS source IDs.}

\subsection{A Grading Scheme for SED Fitting Accuracy}\label{sec:grading}

We utilized this unprecedented, large sample of available spectroscopic classifications for intermediate-mass members of Tr 14 and 16 to benchmark the performance of our SED fitting methodology. For these tests, we included all sources fit with $\chi_0^2/N_{\rm data}\le 4$ (expanding our comparison sample to 154 late-type and 19 early-type stars) and performed a second fitting run in which we incorporated $V$-band photometry from \citet{Hur+12}.

We plotted the $T_{\rm eff}$ model parameter distributions for each source in the Gaia-ESO comparison sample and compared them to the spectroscopically measured \edit1{$T_{\rm eff,S}$} values reported by D17. The underlying probability distributions are typically non-Gaussian. Bimodal parameter distributions are not uncommon, reflecting the degeneracy between a hot photosphere with higher reddening or a cooler star with less reddening.
We defined a set of accuracy grades from A (excellent agreement between SED fitting and spectroscopy) to F (complete failure to agree). Definitions for the accuracy grades (ABCDF) are given in Table~\ref{tab:grades}.
We consider grades ABC to mean the SED modeling successfully \deleted{obtained}\edit1{reproduced} the true $T_{\rm eff,S}$ of the star,\deleted{ albeit} with \edit1{an accuracy of ${\le}10\%$ for A grades, ${\le}30\%$ for BC grades} and decreasing precision from A \edit1{(${\le}5\%$ relative uncertainty)} to C grades \edit1{(${>}10\%$ relative uncertainty)}. DF grades, by contrast, reveal strong or irreconcilable disagreement between the SED models and spectroscopy. D17 could not report spectroscopic $T_{\rm eff,S}$ for early-type stars, but in all cases our SED models also indicated B-type stars, so we consider them the equivalent of AB grades.

These $T_{\rm eff}$ accuracy grades guided us to refine our goodness-of-fit criteria (Section~\ref{sec:fitting}). Among the 154 late-type stars in our expanded spectroscopic comparison sample,
those with best-fit $1<\chi^2_0/N_{\rm data}\le 4$ represented only ${20}\%$ of the sample, indicating a steep decline in fit quality. These relatively poorly-fit sources were dominated by DF grades. We therefore adopted $\chi^2_0 \le N_{\rm data}$ as the robust cutoff for reliable SED fits. \edit1{This cutoff produced a final Gaia-ESO comparison sample of 139 stars. The secondary SED fitting run, which included $V$-band photometry,  reduced the number of well-fit sources by 10\%.
}

\edit1{The distribution of accuracy grades for the IR-only and $V$+IR SED fitting runs are tabulated in the $N_{\rm IR}$ and $N_{\rm V+IR}$ columns, respectively, of Table~\ref{tab:grades}.
  In the IR-only fitting of the full Gaia-ESO comparison sample, 97 sources (70\%) received ABC accuracy grades (this tally includes the 14 early-type stars). Among the well-fit sources in the $V$+IR fitting run, 92 (74\%) received ABC grades. This insignificant gain in the accuracy of the SED fitting in reproducing $T_{\rm eff}$ was produced by the omission of a small number DF sources that were not well-fit when $V$-band photometry was included.
  The inclusion of visual photometry does tend to more tightly constrain the fitting results, as evidenced by the dozen sources that were promoted from C to B grades.
  }

  \edit1{Including $V$-band photometry makes the SED fitting far more sensitive to the shape of the reddening law than is the case with IR photometry alone.
    % , and an inaccurate assumed reddening decreases the fit quality.
    The reddening law is known to vary with location in the CNC, especially in the more obscured regions \citep[e.g.,][]{CCCPOBc} that occupy a large area on the sky outside of the lightly-reddened window studied by \citet{Hur+12} and D17 (see Figure~\ref{fig:overview}). Because the reddening law itself is not a free parameter in our models, any decrease in SED fit quality at visual wavelengths at higher extinction values is unlikely to be outweighed by gains in precision over fitting the IR SED alone. We therefore do not attempt to incorporate visual photometry (for example, from {\em Gaia} DR2) into our subsequent analysis of the full CCCP IR-bright sample.
}

%%%%%%%%%% FIGURE SET: Examples of Individual Grades + AB %%%%%%%%%%%
% 5. SED fit plots
\begin{figure*}[tb]
%%\epsscale{1.}
\centering
\includegraphics[width=0.95\textwidth]{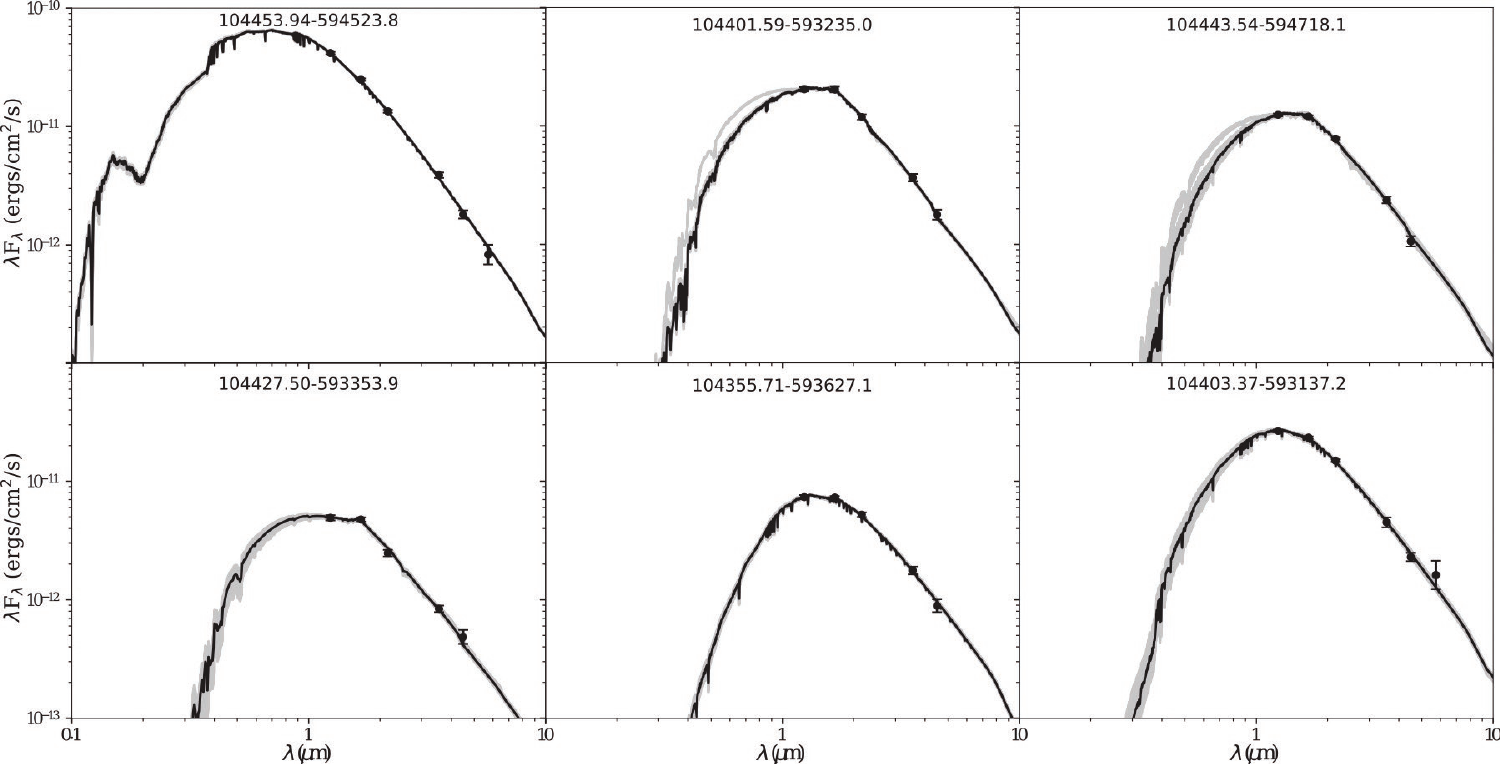}
\\
\caption{\deleted{Example} SED fits \deleted{using our standard,}\edit1{to the 1--8~\um} 2MASS+IRAC IR photometry \deleted{only}\edit1{for six example sources}. In each panel, the \edit1{200} \edit1{best-fitting SED models are} plotted, with the black curve highlighting the \deleted{single best fit}\edit1{model with the minimum $\chi^2$}. \label{fig:IRSEDs} \deleted{The legend gives the CCCP source ID and information about the best-fit model parameters ($\chi^2 = \chi^2_0$). Model names encode $t_{\star}$ (years) and $M_{\star}$ (\Msun).}
}

\vspace{3em}
%\epsscale{0.85}
%\centering
\includegraphics[width=0.8\textwidth]{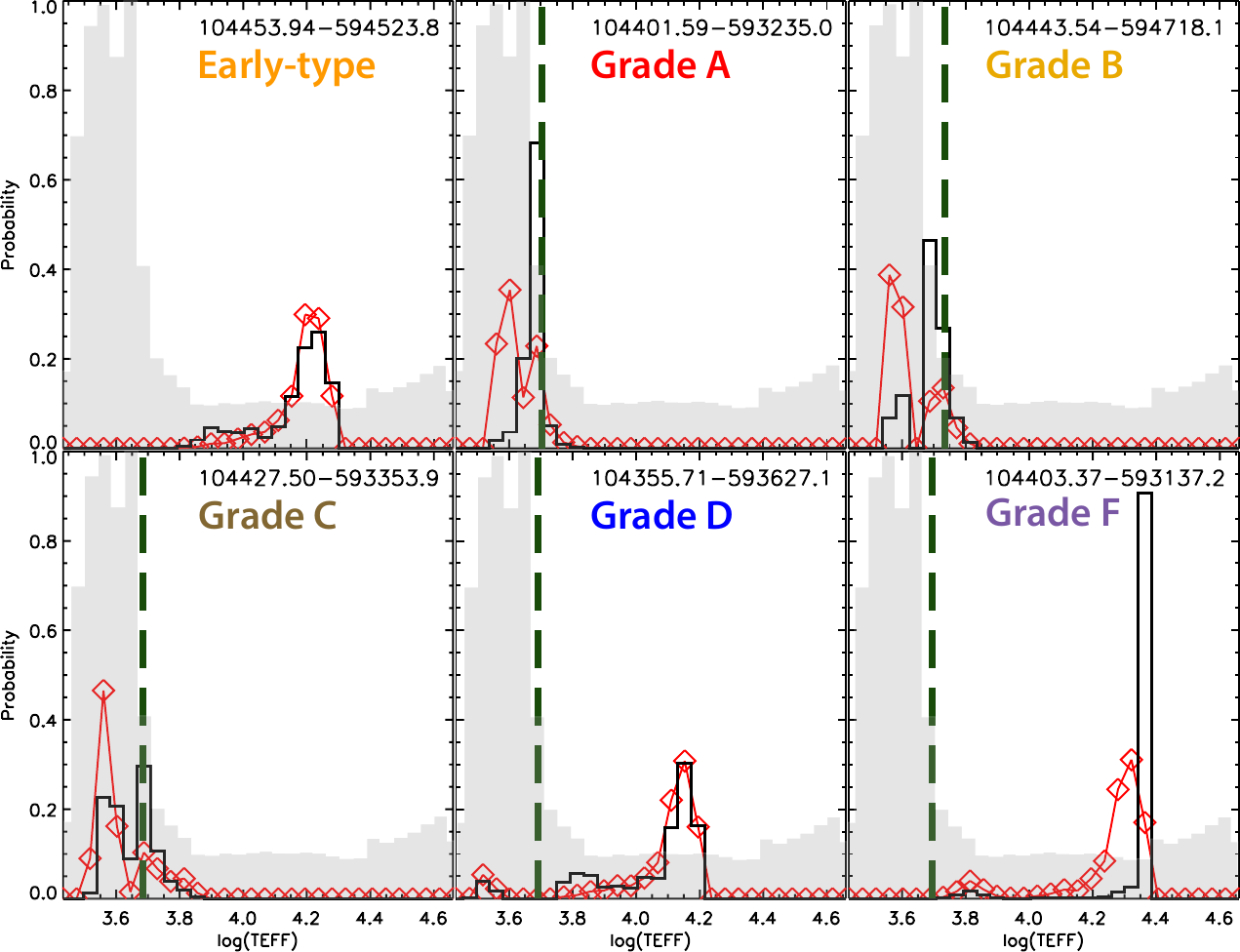}
\caption{Distribution plots of the $T_{\rm eff}$ parameter for the six example SED fits shown in Figure~\ref{fig:IRSEDs}. Each source exemplifies one of the SED accuracy grades described in Table~\ref{tab:grades}. Three histograms are plotted in each panel: the distribution of all models in the naked pre-MS set (purple, same in all panels); the model fitting results weighted by $W_i(\chi^2)$ only (red); and the final model fitting results incorporating all weighting factors in $P_i(\tau_d,\tau_X)$ (black). The thick, vertical long-dashed green lines mark the spectroscopic $T_{\rm eff}$ from D17, the typical uncertainties on which are comparable to the size of one histogram bin in these plots.
\label{fig:Teff_ABCD}
} 
\end{figure*}

\edit1{We chose six example sources to represent each of the five SED accuracy grades plus one early-type star. We plot their SED fits in Figure~\ref{fig:IRSEDs} and present their $T_{\rm eff}$ distributions in Figure~\ref{fig:Teff_ABCD}. The D17 $T_{\rm eff,S}$ values are overplotted as dashed green lines for comparison to the $P_i(\tau_d,\tau_X)$--weighted model parameters, illustrating the basis for our accuracy grade assignments. The SED fitting results weighted {\em only} by goodness-of-fit ($W_i(\chi^2)$; red-diamond curves) fail decisively to predict $T_{\rm eff,S}$ in all of these examples except the early-type star. This is due to the high density of models at cooler temperatures that are inconsistent with the absence of circumstellar disks and envelopes, demonstrating the critical importance of including the $W_i(\tau_d)$ parameter from P16.
}

\subsection{Discrepancies Between SED Fitting and Spectroscopy}
\edit1{
In nearly all cases where we assigned DF accuracy grades (including the two examples shown in Figure~\ref{fig:Teff_ABCD}) the SED fitting results preferred a too-hot $T_{\rm eff}$, equivalent to AB-type stars on the main sequence, while the spectroscopic $T_{\rm eff,S}$ indicated FGK stars. The simplest explanation for these discrepancies is that we have assumed an incorrect distance in the SED fitting, while the spectroscopic $T_{\rm eff}$ measurements are distance-independent. Other potential causes for DF grades include erroneous cross-matches between our sample and D17 (assumed spectroscopic $T_{\rm eff}$ is incorrect), unresolved binaries (or confusion between visual and IR sources in crowded areas), or time-variability across the several years separating the epochs in which the various visual and IR datasets were obtained.}

%%%%%%%%%% FIGURES: EW(Li) and parallax versus Teff, coded by Grade %%%%%%%%%%%
\begin{figure}[thp]
%\epsscale{1.2}
%\centering
\includegraphics[width=\columnwidth]{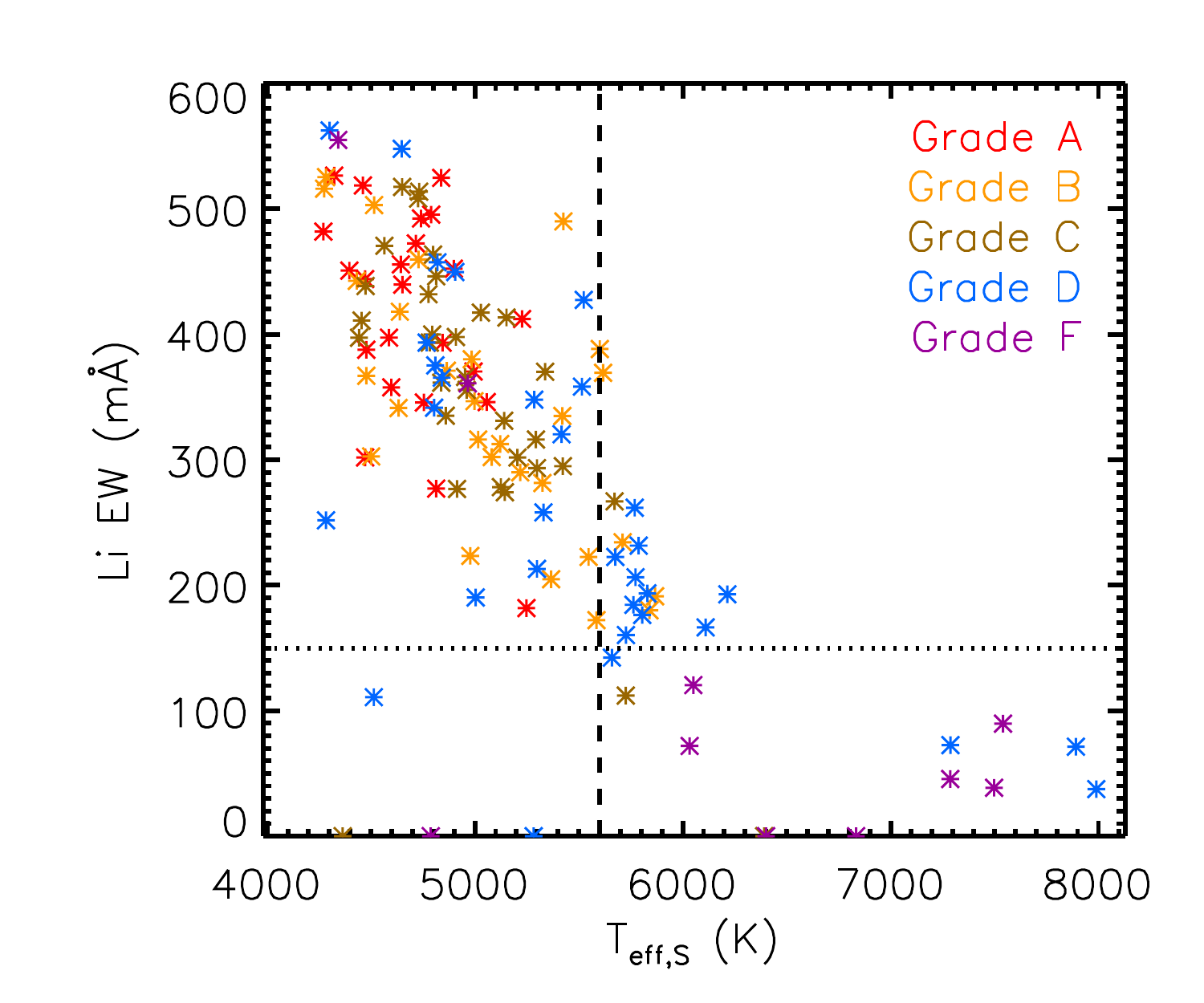}
\includegraphics[width=\columnwidth]{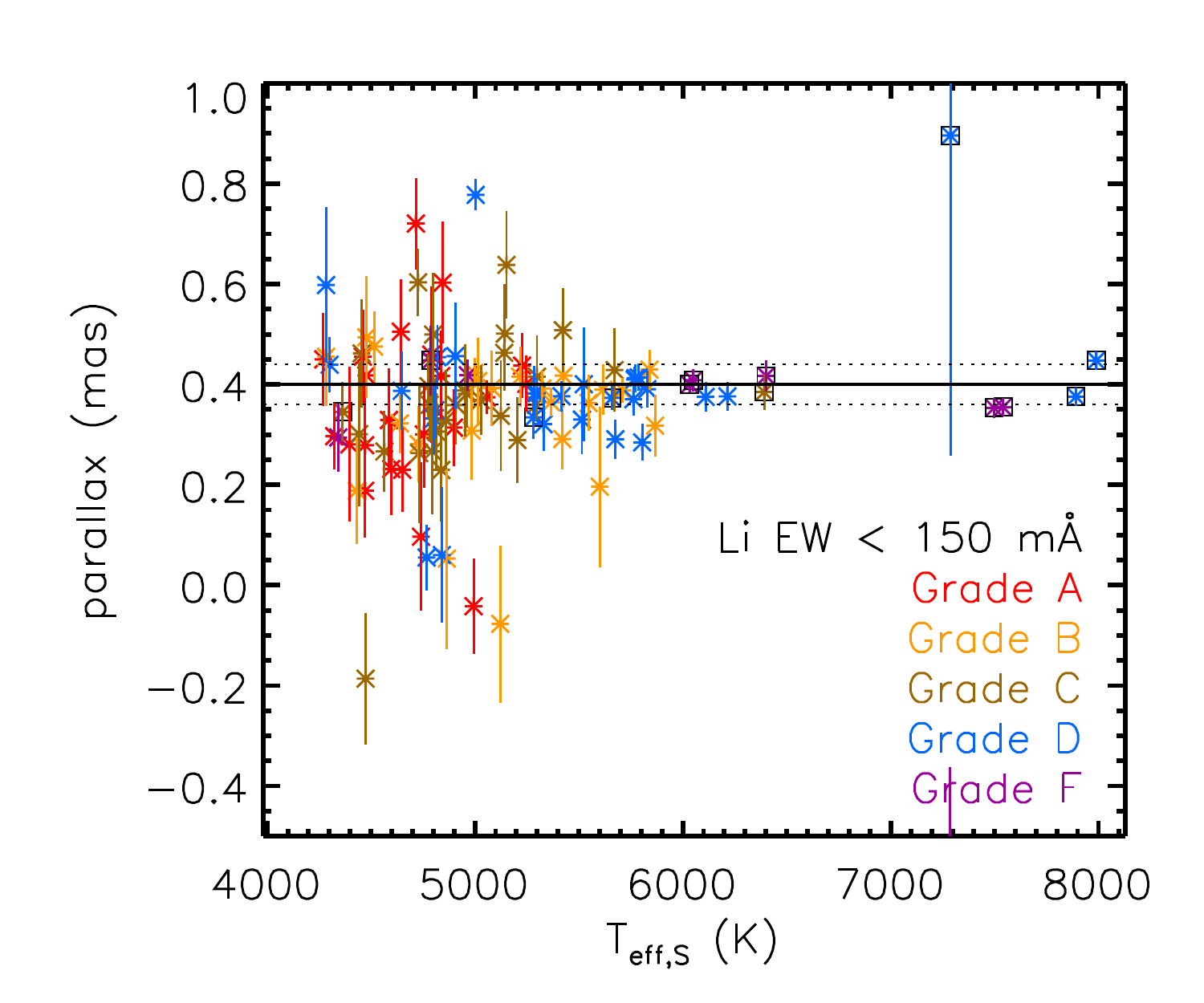}
\caption{{\em Top:} Plot of Li equivalent width versus stellar effective temperature for all stars in the Gaia-ESO cool comparison sample. Symbols are color-coded by their SED-fitting grade. The dotted horizontal line shows the cutoff Li EW$ > 150~$m\AA\ used by D17 for membership selection, \edit1{while the vertical dashed line as $T_{\rm eff,S}=5600$~K provides a rough dividing line between regions dominated by DF and ABC SED accuracy grades.} {\em Bottom:} \edit1{Plot of Gaia DR2 parallax versus stellar effective temperature (12 stars are ommitted because they lack parallax measurements, and one low and one high parallax outlier fall outside the bounds of the plotting range). Black boxes mark sources with Li EW$ < 150~$m\AA. Horizontal solid and dotted lines show the median parallax and uncertainty of $0.40\pm 0.04$ for our sample of CCCP stellar members (Section~\ref{sec:parallax}).}
\label{fig:D17teff}
} 
\end{figure}

\edit1{
In Figure~\ref{fig:D17teff} we plot Li EW (top panel) and {\em Gaia} parallax (bottom panel) versus $T_{\rm eff,S}$ for the late-type stars in the Gaia-ESO comparison sample, color-coded by SED model accuracy grade.
%As D17 showed (their Fig.~5), Li EW tends to decrease with increasing $T_{\rm eff}$.
We find that 13 of 15 stars with Li EW $<150$~m\AA, have DF grades (the other two have C grades). Stars with such low Li EW would have not been selected for membership by D17 if they lacked CCCC X-ray counterparts, increasing the chances that they could be foreground contaminants. However, the parallaxes of the large majority of stars DF grades, including those with Li EW $<150$~m\AA,
exhibit generally smaller error bars and less scatter about the CCCP median value compared to the distribution of sources with ABC grades, strongly suggesting that they are CNC members, not foreground contaminants.}

\edit1{
Stars in the cool comparison sample with $5600~{\rm K}<T_{\rm eff,S}<8000$~K are dominated by DF grades (Figure~\ref{fig:D17teff}), but their lower Li EW is consistent with their earlier spectral types (A6 through G5; \citealp{PM13}), and it is not necessarily indicative of the older ages expected of field contaminants. These stars are also visually brighter than typical of the comparison sample, which may explain their relatively low {\em Gaia} parallax uncertainties. They have $\log{L_{\rm bol}/L_{\sun}}\ga 1$, placing them in a region of the HR diagram where the density of the naked stellar models is minimal (Figure~\ref{fig:params}, top panel). The SEDs of stars in this region of parameter space are also well-fit by ZAMS models at just slightly higher $A_V$. Because the model density near the ZAMS is very high, the distribution of well-fit models becomes incorrectly skewed toward higher $T_{\rm eff}$ values. We suspect this effect is responsible for the majority of the DF grades in the Gaia-ESO comparison sample, and we discuss its impact on our results in the next section.}

\section{Results: Probabilistic H-R Diagrams}
Thus far we have examined only the SED fitting results for individual sources. However, we are primarily interested in the aggregate properties of stellar populations. We obtain these by summing the normalized SED model parameter distributions of all constituent sources within the various subsets of interest.

%%%%%%%%% FIGURE: Teff Distribution Function for Comparison Sample %%%%%%%%%%%
% 2 Panels -- Without and With V-band photometry incorporated

%\begin{figure*}[thp]
\begin{figure}[tpb]
%\epsscale{1.1}
%\centering
\includegraphics[width=\columnwidth]{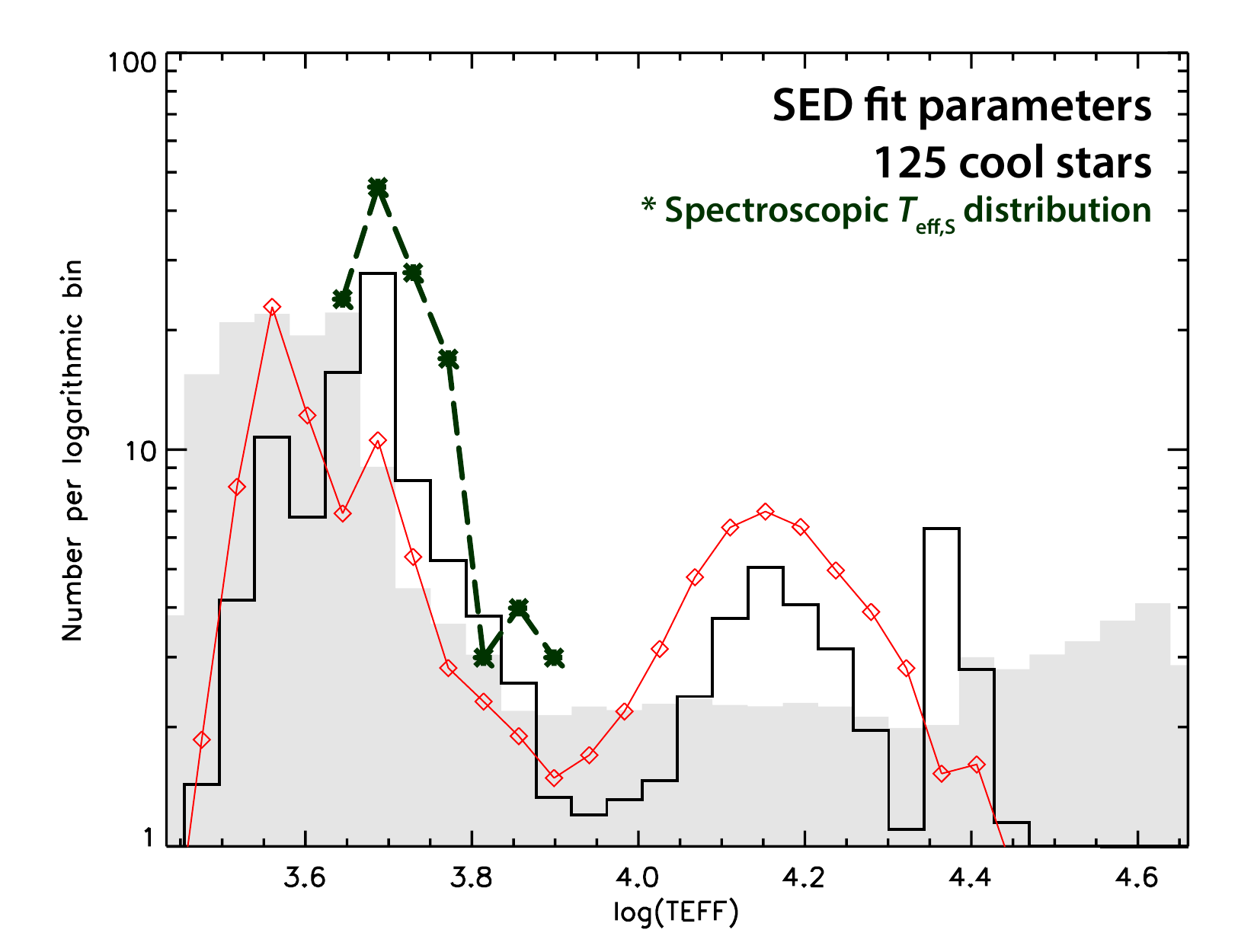}
\caption{Combined $T_{\rm eff}$ distributions for all SED fits to the cool spectroscopic comparison sample. Colors and line/histogram styles are the same as in Figure~\ref{fig:Teff_ABCD}.
\label{fig:Teff_Func}
} 
%\end{figure*}
\end{figure}

\subsection{Temperature and Age Distributions for the Spectroscopic Comparison Sample}\label{sec:pHRDcomps}

The aggregate $T_{\rm eff}$ parameter distribution for the late-type stars in the cleaned spectroscopic comparison sample is provided in Figure~\ref{fig:Teff_Func}. This plot shows that our weighted SED fitting parameters (black histogram) successfully reproduce the peak in the spectroscopic temperature distribution near $\log{T_{\rm eff}/{\rm [K]}}=3.7$. \edit1{The $T_{\rm eff}$ distribution weighted only by $\chi^2$ (red histogram) is unsuccessful, producing a large peak at cooler temperatures that would indicate disk-bearing stars, and a smaller peak at the correct temperature. Irrespective of the weighting function used, SED fitting produces a spurious hump in the distribution with $\log{T_{\rm eff}/{\rm [K]}} > 4$, corresponding to the 30\% of sources with DF accuracy grades.}
\deleted{Including $V$-band photometry would not improve the agreement between SED fitting and spectroscopy, hence we restrict our subsequent discussion of the Gaia-ESO comparison sample to the IR-only SED fitting results.}

%%%%%%%%% FIGURE: pHRDs for Comparison Sample %%%%%%%%%%%
% 2 Panels -- Regular, and using Teff as parameter constraint. SKIPPING the V-band stuff as it's not helpful, adds NOTHING if you happen to have spectra.
\begin{figure}[tbp]
%%\epsscale{1.0}
\centering
\begin{overpic}[width=0.9\columnwidth]{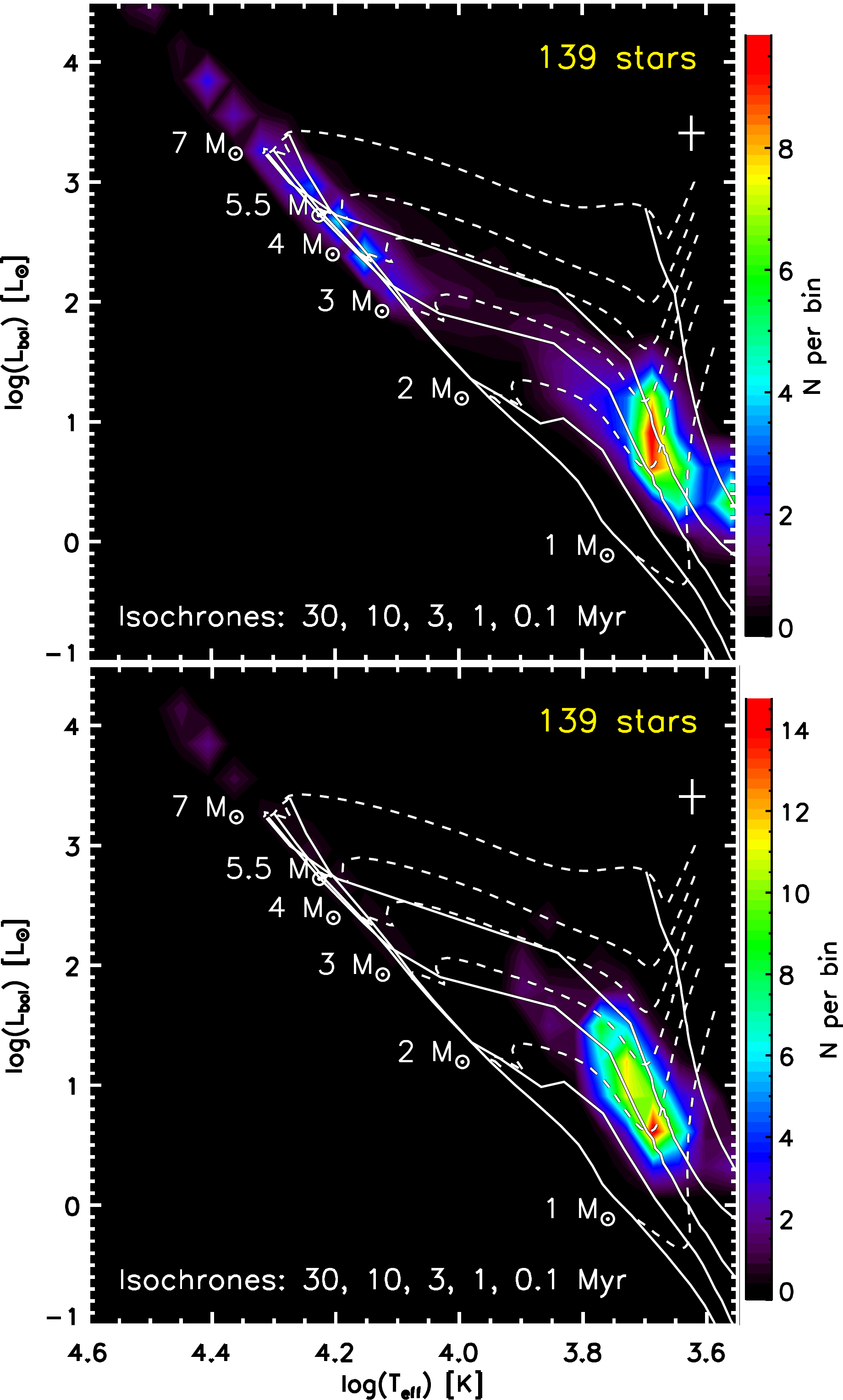}
  \put (10,60) {\color{yellow}\large$P_i(\tau_d,\tau_X)$}
  \put (10,12) {\color{yellow}\large$P_i(T_{\rm eff})$}
\end{overpic}
\caption{Probabilistic H--R diagrams for all SED fits to the spectroscopic comparison sample. \citet{SDF00} isochrones and evolutionary tracks are plotted as solid and dashed curves, respectively, for the indicated stellar masses and ages. The white crosses in the upper-right give the 2-D histogram binsizes. {\em Top:} Results from fitting IR-only SEDs using our standard $P_i(\tau_d,\tau_X)$ weighting function. {\em Bottom:} Same SED fits, but using the $P_i(T_{\rm eff})$ weighting function to constrain the fitting results for the \edit1{125} cool stars using spectroscopic data from D17 (the 14 early-type stars without $T_{\rm eff}$ measurements are weighted only by $W_i(\chi^2)$).
\label{fig:pHRDs_D17}
} 
\end{figure}

We also construct joint probability distributions of $T_{\rm eff}$ and $L_{\rm bol}$ and plot them directly on the traditional H--R diagram as two-dimensional histograms.
Summing the probability distributions for all of the individual stars in the spectroscopic comparison sample, we obtain the {\it probabilistic H--R diagram} (pHRD; Figure~\ref{fig:pHRDs_D17}).\deleted{\footnote{We omit analogous pHRDs for SED fits including $V$-band photometry because they do not appear noticeably different.}} pHRDs give similar information to a traditional color-magnitude diagram, and indeed they appear qualitatively similar to the distribution of points on the $V_0$ versus $(V-I)_0$ CMD presented by D17 for the entire Gaia-ESO sample (their Fig.~13), of which our comparison sample is a subset.
The pHRD represents a philosophical shift away from using a single color and magnitude for each individual star to infer the ensemble properties of a larger population. Instead, we synthesize a model for the ensemble population, using a multidimensional set of colors and magnitudes for each constituent star.

In Figure~\ref{fig:pHRDs_D17} we compare pHRDs produced using two different weighting methods for the SED fitting results: ({\em top} panel) our standard, age-based likelihood weighting (\edit1{$P_{i}(\tau_d,\tau_X)$}; Section~\ref{sec:weighting}) and ({\em bottom} panel) a Gaussian weighting factor \edit1{$W_i(T_{\rm eff,S})$} using \edit1{$T_{\rm eff,S}$} and its uncertainty reported by D17. The  weighting function in the latter case is
\begin{equation}\label{eq:PiT}
  P_i(T_{\rm eff,S}) = P_nW_i(\chi^2)W_i(T_{\rm eff,S}),
\end{equation}
which typically (and unsurprisingly) yields much tighter constraints on all model parameters for individual sources.
\edit1{While both pHRDs place their pre-MS loci between the 1 and 3 Myr isochrones, the shapes of these loci are noticeably different between the two plots. The peak probability in the $P_{i}(\tau_d,\tau_X)$-weighted case forms a narrow ridge with $3.65<\log{T_{\rm eff}/{\rm [K]}}<3.7$, which extends in the $\log{L_{\rm bol}}$ direction upwards from the 2~\Msun\ evolutionary track to nearly the 3~\Msun\ track (red part of the heat map in the left panel, with ${>}8$ stars per bin). The pre-MS locus in $P_i(T_{\rm eff,S})$-constrained case, by contrast, extends parallel to the isochrones, and its peak is much sharper, located on the 2~\Msun\ track just above the 3~Myr isochrone (right panel, ${\sim}14$ stars per bin).} Two other features of the pHRD produced using our likelihood weighting (Figure~\ref{fig:pHRDs_D17}, {\em top} panel) are missing from the $T_{\rm eff,S}$-constrained pHRD ({\em bottom} panel): (1) a residual locus of cool ($\log{(T_{\rm eff}/{\rm [K]})}<3.6$), low-mass ($M_{\star}<1$~\Msun) pre-MS stars and (2) the intermediate-mass (3--7~\Msun) ZAMS. \edit1{Our standard likelihood procedure tends to relocated with DF grades relocated from the hot end of the pre-MS locus ($\log{T_{\rm eff,S}/{\rm [K]}}\approx 3.75$) to the ZAMS (see also Figure~\ref{fig:D17teff}). Because this displacement generally parallels the isochrones, the age determination of the population is unaffected, but the masses of stars with DF grades are overestimated. }

%%%%%%%%% FIGURES: Mass-age Histograms for Comparison Sample %%%%%%%%%%%
\begin{figure}[tb]
%%\epsscale{1.0}
\centering
\begin{overpic}[width=0.9\columnwidth]{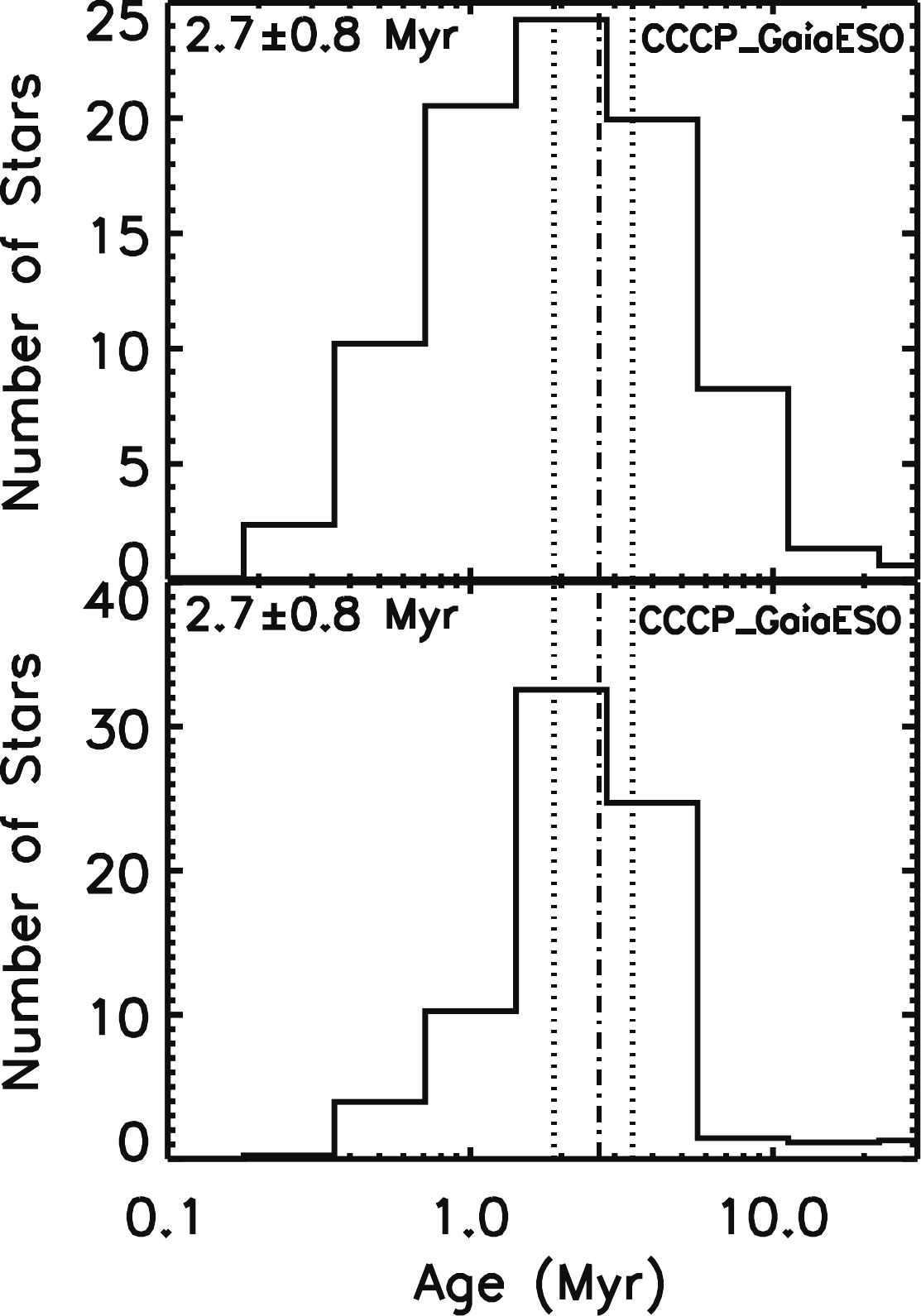}
  \put (52,92) {\large$P_i(\tau_d,\tau_X)$}
  \put (52,47) {\large$P_i(T_{\rm eff,S})$}
\end{overpic}
\caption{Age ($t_{\star}$) distributions for all SED models fit to the spectroscopic comparison sample. The panels correspond to the pHRDs shown in Figure~\ref{fig:pHRDs_D17}: standard weighting ({\em top}) and spectroscopic temperature constraints {\em bottom}.\deleted{: ($P_i(\tau_d,\tau_X)$ weights {\em left}, $P_i(T_{\rm eff,S})$ constraints on the {\em right}).}
In each plot, $\tau_{\rm SF}$ and its uncertainty are annotated and indicated by vertical dash-dotted and dashed lines.
\label{fig:pAge_D17}
} 
\end{figure}
\begin{figure}[thp]
%%\epsscale{0.8}
\centering
\begin{overpic}[width=0.8\columnwidth]{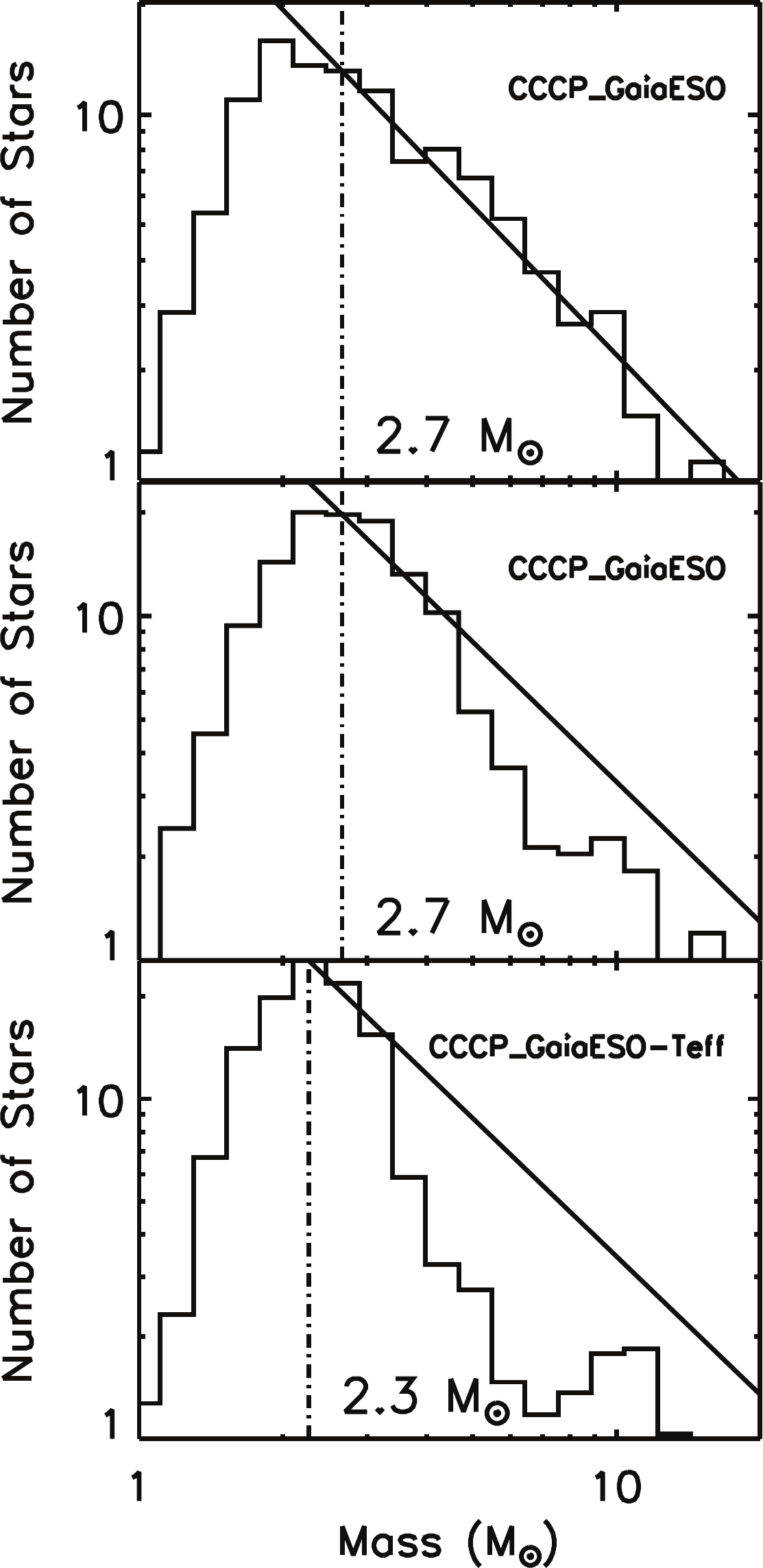}
  \put (32,90) {\large$P_i(\tau_d,\tau_X)$}
  \put (30,59.5) {\large$P_i(\tau_d,\tau_X,\tau_{\rm SF})$}
  \put (32,28.5) {\large$P_i(T_{\rm eff,S})$}
\end{overpic}
\caption{Mass ($M_{\star}$) distributions for all SED models fit to the spectroscopic comparison sample \edit1{using three different weighting options. {\em Top:} Standard weighting (corresponding to the top panels of Figures~\ref{fig:pHRDs_D17} and \ref{fig:pAge_D17}).
{\em Center:} Isochronal weighting (Equation~\ref{eq:PiSF}) using $\tau_{\rm SF}=2.7\pm 0.8$~Myr.
{\em Bottom:} spectroscopic temperature constraints (corresponding to the bottom panels of Figures~\ref{fig:pHRDs_D17} and \ref{fig:pAge_D17}).} In each plot, solid lines show the Salpeter power-law slope scaled to the distribution at the cutoff mass $M_{C}$ (annotated and indicated by vertical dash-dotted lines).
\label{fig:pMass_D17}
} 
\end{figure}
Because pre-MS tracks and isochrones are built into the SED models, we can trivially transform the pHRDs (Figure~\ref{fig:pHRDs_D17}) into probability distributions of  $t_{\star}$ and $M_{\star}$  (Figures~\ref{fig:pAge_D17} and \ref{fig:pMass_D17}).  Compared to the pHRD (or the traditional HRD), these transformations offer more straightforward visualizations of the fundamental stellar parameter distributions. To construct these distributions, we bin the well-fit models uniformly in logarithmic intervals, sampling across the full range of each parameter dimension (Figure~\ref{fig:params}) and applying the $P_i(\tau_d,\tau_X)$ weighting functions described in Section~\ref{sec:weighting} to each source. We experimented with different numbers of bins for each parameter and chose an optimal binning for the composite histograms that provided the smallest binsize for which fluctuations between adjacent bins did not appear dominated by noise.
The final binnings chosen for the 2D pHRDs and associated 1D-distributions of $M_{\star}$ and $t_{\star}$ were based on the total number of sources included in each distribution.

We create and analyze age and mass distributions (except for the  $P_i(T_{\rm eff,S})$-constrained models) using a two-step, iterative process. In the first iteration, we identify the modal bin where $M_{\star}=M_{C1}$ in the mass distribution produced with our standard $P_i(\tau_d,\tau_X)$ likelihood weighting. We include only those models with \edit1{$M_{\star}\ge 1.8$~\Msun\ } in the $t_{\star}$ distributions (Figure~\ref{fig:pAge_D17}). We adopted this cutoff mass to avoid biasing the age distributions \edit1{toward the youngest pre-MS stars. \edit1{It also removes the spurious locus of too-cool models with $\log{T_{\rm eff}/{\rm [K]}}<3.6$ (Figures~\ref{fig:D17teff} and \ref{fig:pHRDs_D17}).} The relatively shallow 2MASS and Vela-Carina surveys are insensitive to ZAMS stars fainter than early F dwarfs at the distance of the CNC}. Younger, low-mass pre-MS stars are larger and and cooler, hence they may be detected and included in our sample while older stars of the same mass fall below our IR photometric sensitivity limits. \edit1{The existence of fainter, low-mass stars can be inferred from the age distributions of intermediate-mass stars.}
%For stars on the main sequence, only a lower age limit corresponding to the ZAMS arrival timescale, $\tau_{\rm ZAMS}(M_{\star})$, is meaningful. Our implementation of the $W_i(\tau_X)$ weighting function (Equation~\ref{eq:xweight}) ensures that models representing intermediate-mass stars on the ZAMS prefer this minimal age.

In the case of a star-forming event that commenced at time $\tau_{\rm SF}$ (Myr) in the past and proceeded with approximately constant star-formation rate (SFR) to the present day, logarithmically-binned age distributions should exhibit a linearly-increasing number of stars per bin as $t_{\star}$ increases from 0 to $\tau_{\rm SF}$, with a sharp break in the distribution for $t_{\star}>\tau_{\rm SF}$.
To locate this break point in our age distributions, we identify the modal bin in the $t_{\star}$ histogram (Figure~\ref{fig:pAge_D17}). The duration of star formation $\tau_{\rm SF}$ is the geometric mean of the central values for this modal bin and the adjacent bin in the direction of increasing $t_{\star}$.
The uncertainty on $\tau_{\rm SF}$ is the geometric mean of the widths of these two bins. 

Both $t_{\star}$ distributions for the comparison sample give the same value for $t_{\rm SF}=2.7\pm 0.8$~Myr (Figure~\ref{fig:pAge_D17}). While spectroscopy sharpens the age distribution, our likelihood-weighted SED fitting to X-ray selected, diskless sources accurately measures the duration of star formation from broadband IR photometry alone.

\edit1{Three model} mass distributions for the spectroscopic comparison sample are plotted in Figure~\ref{fig:pMass_D17}.
\edit1{The distribution created with our standard weighting function ({\em top} panel) follows a Salpeter power-law for $M_{\star}\ge 2.7~M_{\sun} = M_C$, turning over due to photometric incompleteness at lower masses. However, the $P_i(T_{\rm eff,S})$-constrained mass distribution, which is both more precise and more accurate, appears strikingly different. The peak of the distribution narrows, shifting to a marginally lower $M_C=2.3$~\Msun\ and producing a significant deficit of stars with $M_{\star}>3$~\Msun\ when compared to the Salpeter slope (the area between the solid line and histogram in the {\em bottom} panel).}
Sample bias introduced by our X-ray selection criteria explains this observed deficit.
\edit1{Intermediate-mass (2--8~\Msun) stars of types A and late B on the MS have no known mechanism for producing detectable X-ray emission and hence occupy an ``X-ray desert'' between cooler stars with coronal emission \citep{P05} and hotter OB stars producing X-rays via shocks driven by strong winds \citep{G11}.
  But those with X-ray bright T Tauri companions \citep[e.g.,][]{CCCPcomps} can be selected for inclusion in our IR-bright, diskless sample.}

\edit1{To better constrain the model mass distribution in the general case where no $T_{\rm eff,S}$ is available, we iterated the analysis of the SED fitting results,} introducing a new Gaussian weighting parameter $W_i(\tau_{\rm SF})$. The complete weighting function for this second iteration is hence
\begin{equation}~\label{eq:PiSF}
  P_i(\tau_d,\tau_X,\tau_{\rm SF}) = P_nW_i(\chi^2)W_i(\tau_d)W_i(\tau_X)W_i(\tau_{\rm SF}).
\end{equation}
For determining the present-day mass function of a stellar population from broadband photometry, it is a common practice to construct a dereddened luminosity function and subsequently to employ a mass--luminosity relation appropriate to the age of the population \citep{O+14_IMF}.
Here we have adopted $\tau_{\rm SF}$ from the first iteration as the best isochrone for converting from $L_{\rm bol}$ to $M_{\star}$ for our ensemble population\edit1{, producing the mass distribution plotted in the {\em center} panel of Figure~\ref{fig:pMass_D17}}.

\edit1{The two most important metrics for the modeled $M_{\star}$ distributions are the location ($M_C$) and normalization of the scaled Salpeter-type IMF. Scaling a standard IMF model \citep{KW03} to the isochronally-weighted distribution predicts the same number of intermediate-mass stars as scaling to the spectroscopically-constrained distribution (Figure~\ref{fig:pMass_D17}, {\em center} versus {\em bottom} panels) to within 5\% ($N_{\rm IM}\approx 150$ for $M_{\star} > 1.8$~\Msun; Table~\ref{tab:results}). The same IMF scaled to the mass distribution produced using $P_i(\tau_d,\tau_X)$ weights alone predicts ${\sim}30\%$ fewer stars. Poorer constraints on the mass of individual stars broadens the aggregate mass distribution, mimicking a shallower IMF slope and reducing the peak height near the cutoff mass $M_C$  (Figure~\ref{fig:pMass_D17}, {\em top}). We therefore choose to employ the second iteration to leverage isochronal weighting for $M_{\star}$ distributions for the full CCCP IR-bright, diskless source sample.}

\edit1{We caution that our omission of disk-bearing YSOs and X-ray quiet stars makes this sample formally incomplete at all masses. These mass distributions should not be used to make measurements that require completeness over some mass interval, for example circumstellar disk fractions or the total size of the CNC stellar population. They are mainly useful for drawing comparisons among the various CCCP sub-populations.}

\subsection{Age and Mass Distributions for the CCCP Sub-Regions}

Having validated our methodology on the spectroscopic comparison sample, we applied the same analysis to each of our five CCCP spatial sub-regions (Figure~\ref{fig:overview}). In Figures~\ref{fig:RegA1}--{\ref{fig:RegD} we present the pHRDs, age, and mass distributions for the sources in each sub-region. Each pHRD is annotated with the total number $N$ of stars in the sample, and \citet{SDF00} isochrones and evolutionary tracks are overlaid (same as in Figure~\ref{fig:pHRDs_D17}). Vertical lines overplotted on the age and mass distributions give the duration of star formation $\tau_{\rm SF}$ and break mass $M_{C}$, respectively (as in Figures~\ref{fig:pAge_D17} and \ref{fig:pMass_D17}). These and other important quantities describing each sub-region are listed in Table~\ref{tab:results}.

\begin{figure}[tbp]
%%\epsscale{1.}
\centering
\includegraphics[width=\columnwidth]{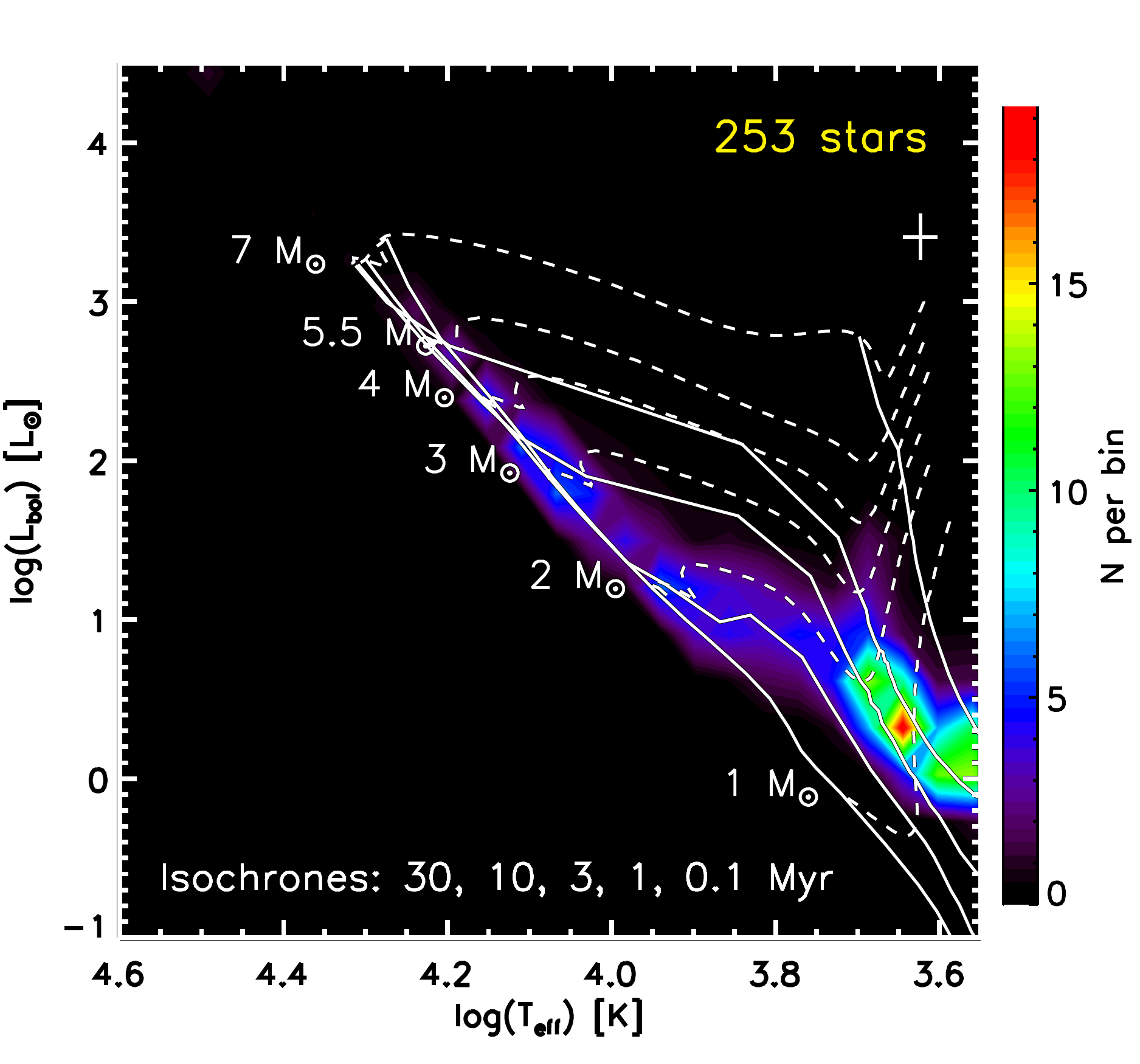}
\includegraphics[width=0.8\columnwidth]{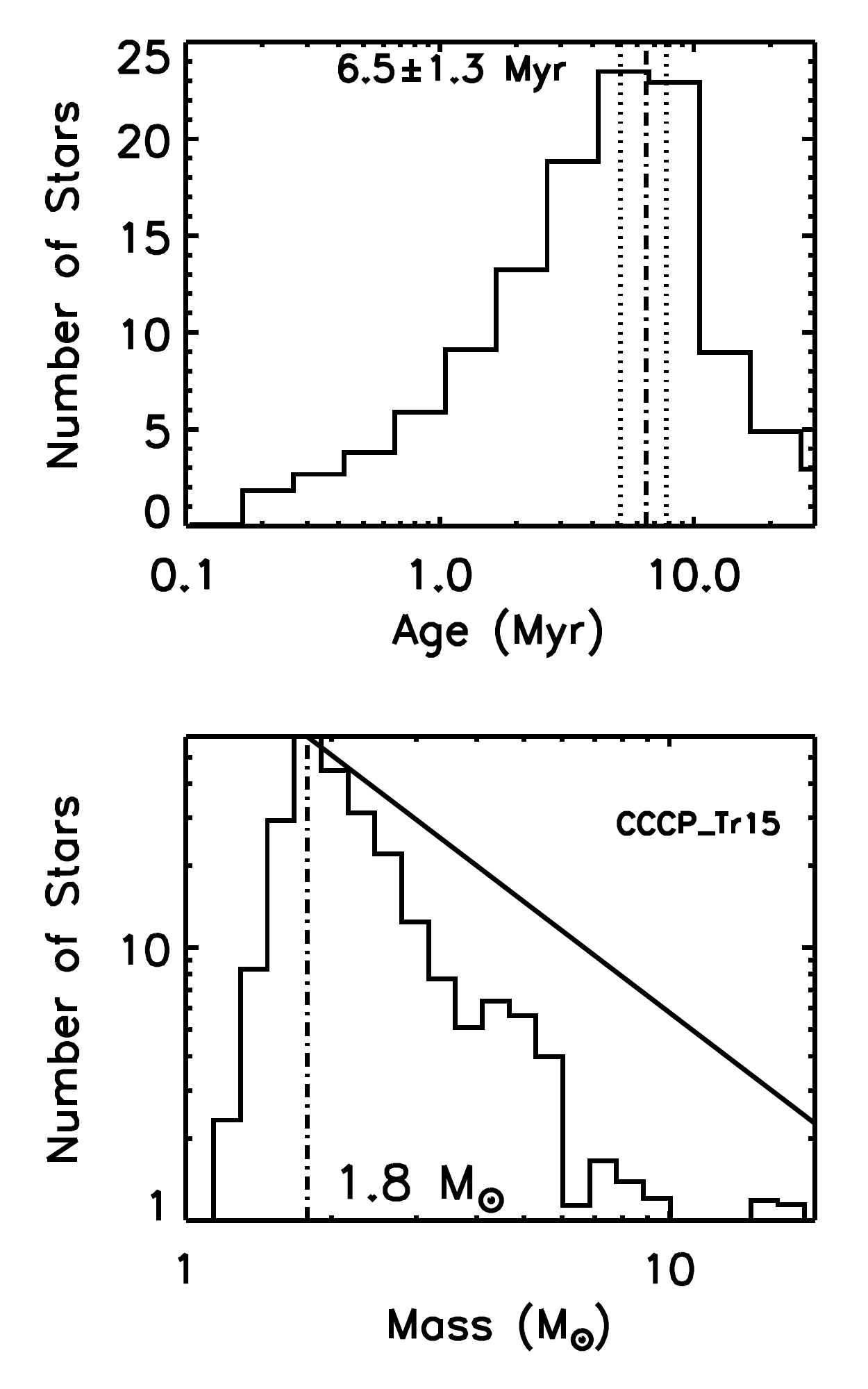}
\caption{{\em Top to bottom:} pHRD, $t_{\star}$ and $M_{\star}$ distributions for CCCP Region A1: Tr 15 and environs. 
\label{fig:RegA1}
}
\end{figure}
Previous studies have found, based on multiple lines of evidence, that Tr 15 is significantly older than Tr 14 and 16, the other two massive clusters in the CNC \citep{Tapia+03,CCCP_Tr15}.
%Some authors have suggested that Tr 15 is a background cluster, based on differences in its age and visual distance modulus compared to the other two Tr clusters [CITATIONS]. {\em Gaia} parallaxes and proper motions leave no doubt that Tr 15 is part of the CNC stellar population \citep{K19_Gaia}.
We confirm a significantly earlier onset of star formation in Tr 15 compared to most of the rest of the CNC, with $\tau_{\rm SF}=6.5\pm1.3$~Myr, in good agreement with the measurements reported by \citet{FFM80_Tr15} from $UBVRI$ photometry and spectroscopy of its brightest members. Tr 15 does not contain a large population IMPS, as most stars with $M_{\star}>2$~\Msun have already reached the ZAMS (Figure~\ref{fig:RegA1}). \edit1{Disk-bearing, bright YSOs are absent from Tr 15 \citep{CCCPYSOs}.}

\begin{figure}[tbp]
%%\epsscale{1.}
\centering
\includegraphics[width=\columnwidth]{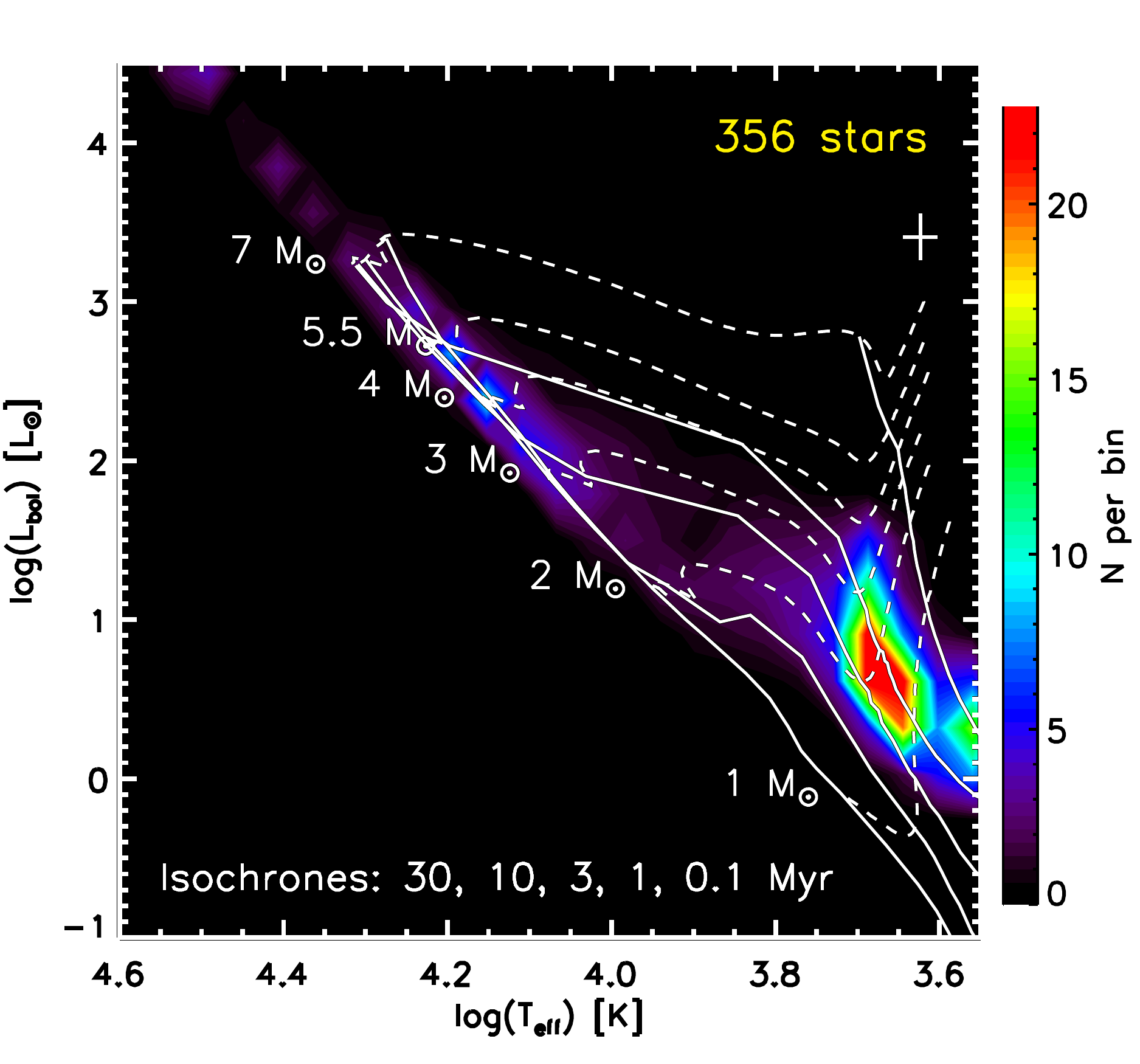}
\includegraphics[width=0.8\columnwidth]{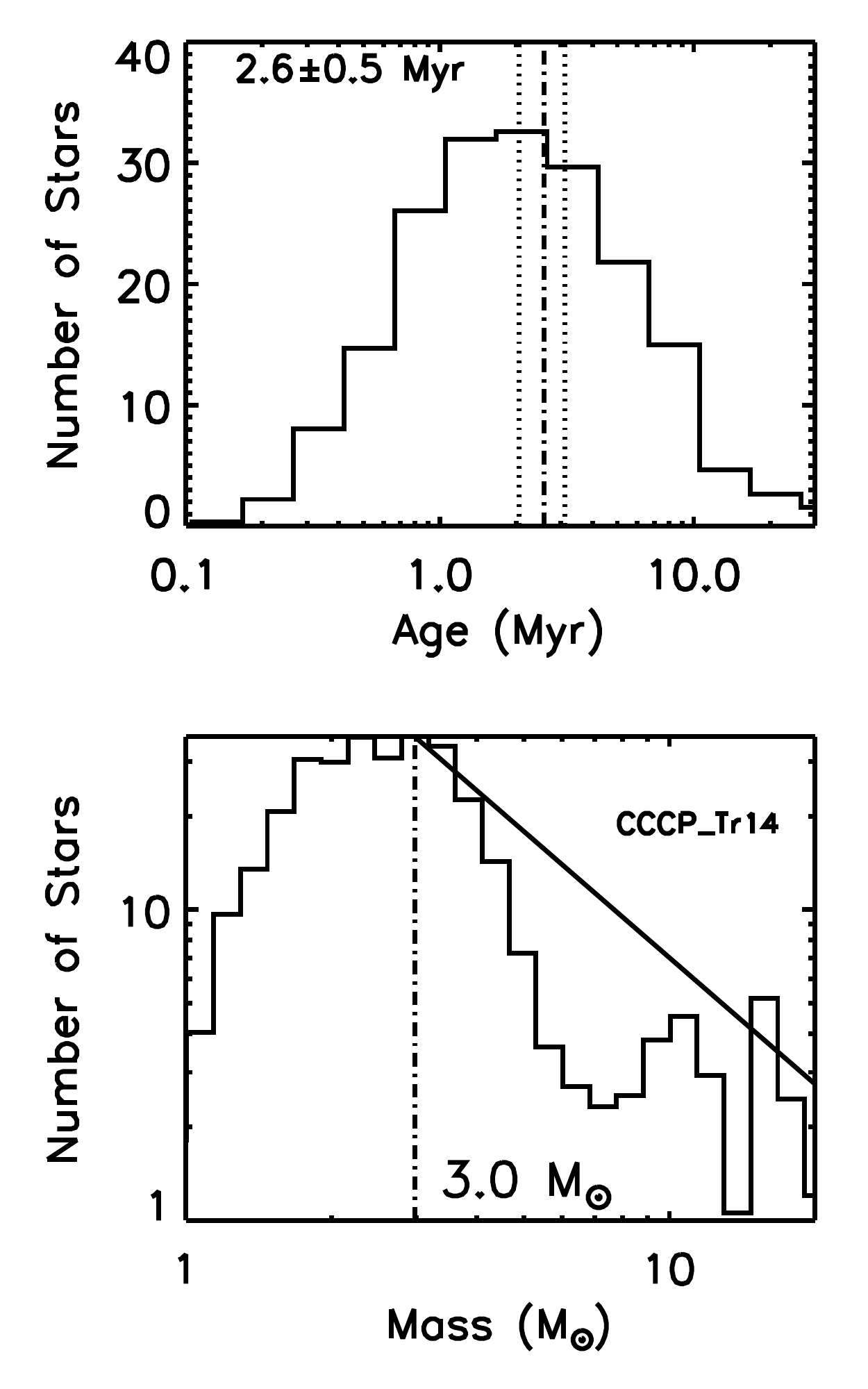}
\caption{{\em Top to bottom:} pHRD, $t_{\star}$ and $M_{\star}$ distributions for CCCP Region A2: Tr 14, Collinder 232, and environs.
\label{fig:RegA2}
}
\end{figure}

\begin{figure}[tbp]
\centering
\includegraphics[width=\columnwidth]{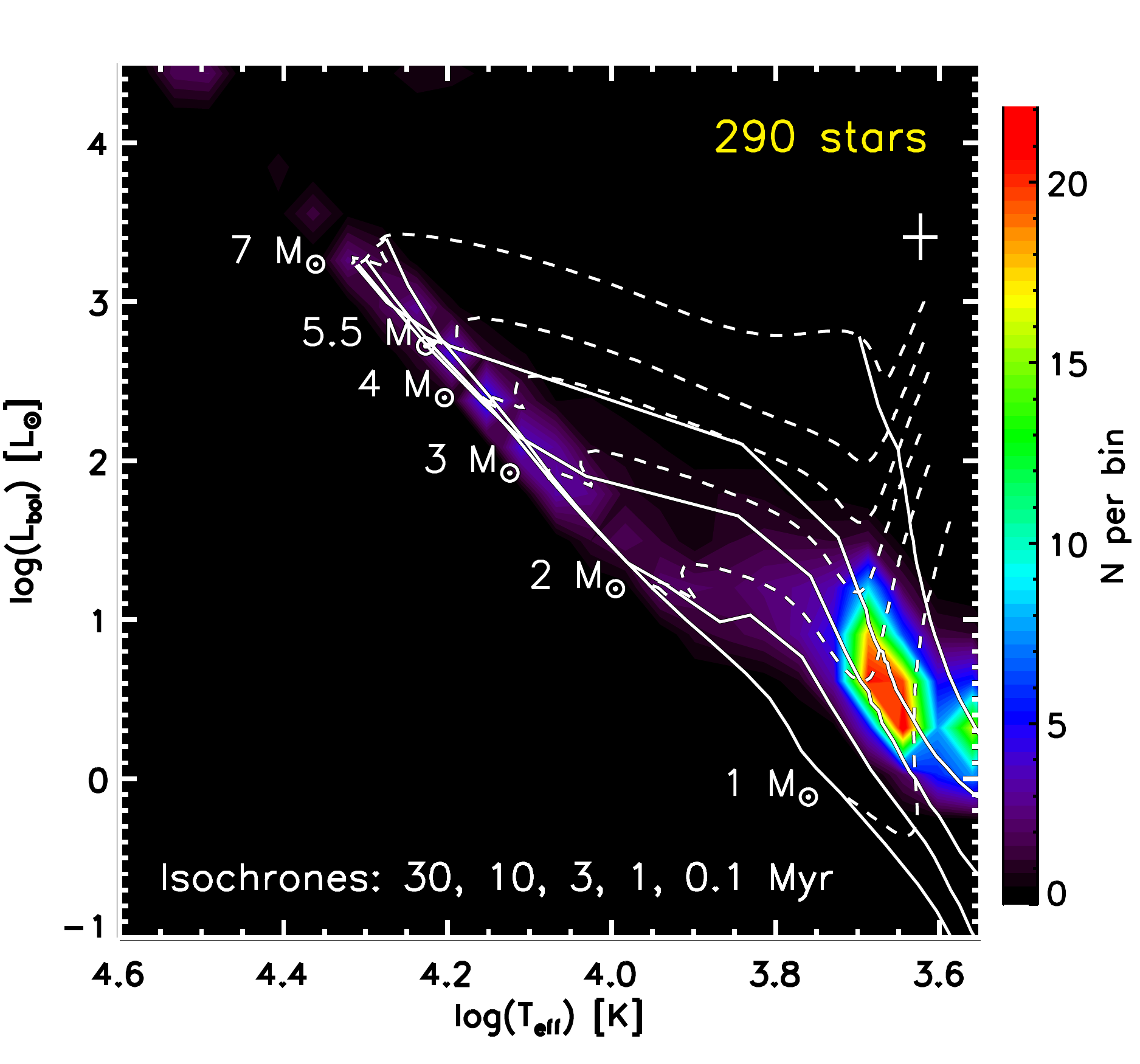}
\includegraphics[width=0.8\columnwidth]{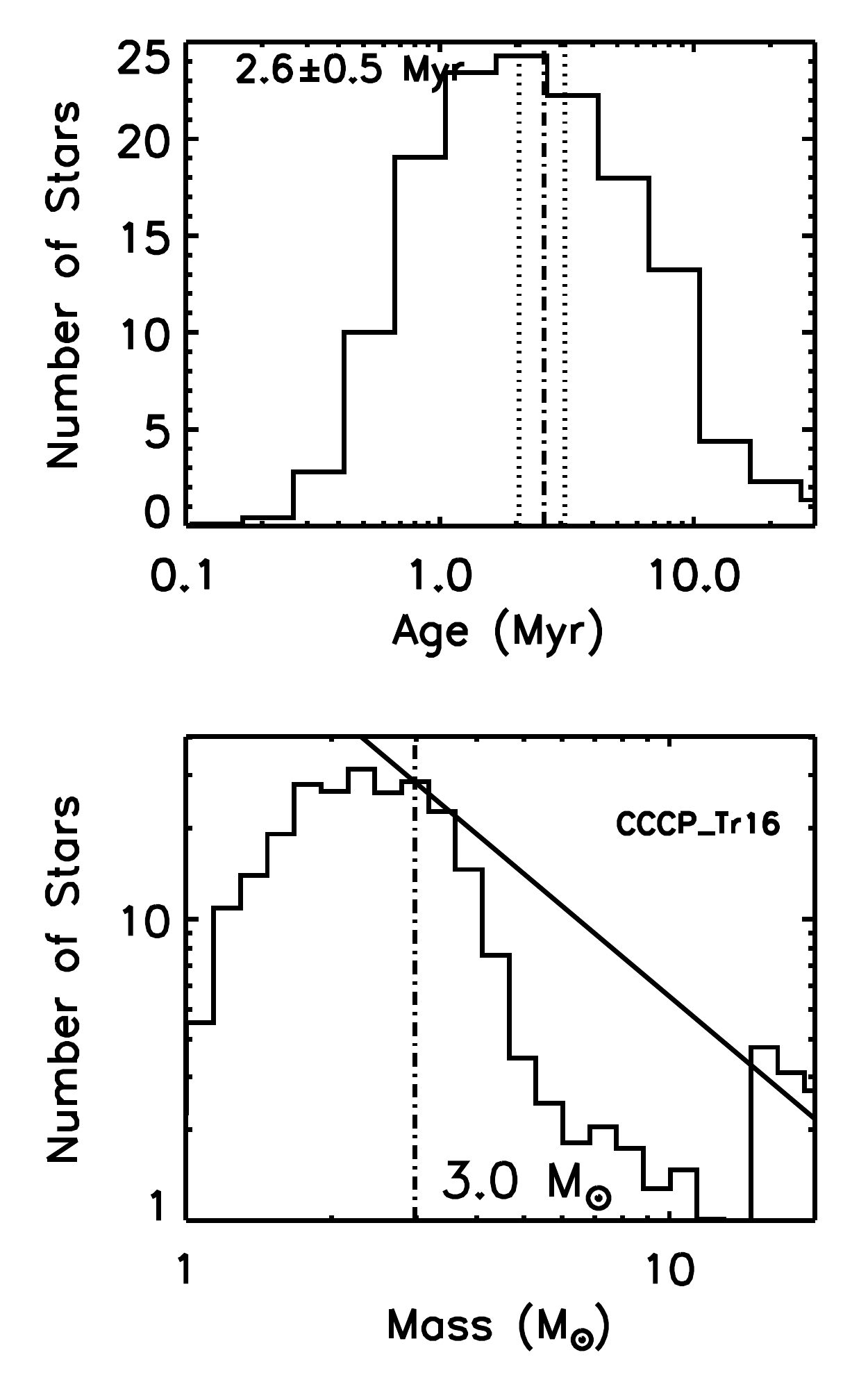}
\caption{{\em Top to bottom:} pHRD, $t_{\star}$ and $M_{\star}$ distributions for CCCP Region B, Tr 16 and environs.
  \label{fig:RegB}
}
\end{figure}
We find $\tau_{\rm SF}=2.6\pm 0.5$~Myr for sub-regions A2 and B, containing Tr 14 and Tr16, respectively, each presenting a very luminous pre-MS populated by 2--3~\Msun\ IMPS (Figures~\ref{fig:RegA2} and \ref{fig:RegB}). Our results agree broadly with previous studies (e.g, \citealp{Tapia+03,A+07_Tr14,Hur+12}). Several studies have reported even younger ages ${<}2$~Myr for the dense core of Tr 14 (\citealp{A+07_Tr14, AgeJX}; D17), and we do not dispute these results. A very young age is consistent with the large concentration of luminous YSOs \citep{CCCPYSOs} and X-ray bright IMPS (Nu\~{n}ez et al.\ 2019, in preparation) in Tr 14. Because we selected diskless stars, our sample was biased (by design) toward more-evolved objects. The core of Tr 14 is confused in the 2MASS and {\em Spitzer}/IRAC images, so our present analysis can only probe the outskirts of the cluster. Our $M_{\star}$ distributions for both sub-regions A2 and B break from the Salpeter slope below a relatively high $M_{C}=3.0$~\Msun, indicating severe photometric incompleteness caused by the combination of crowding (Tr 14), proximity to the very bright IR source $\eta$~Car (Tr 16), and bright, diffuse background IR nebulosity (both).

\begin{figure}[thb]
\centering
\includegraphics[width=\columnwidth]{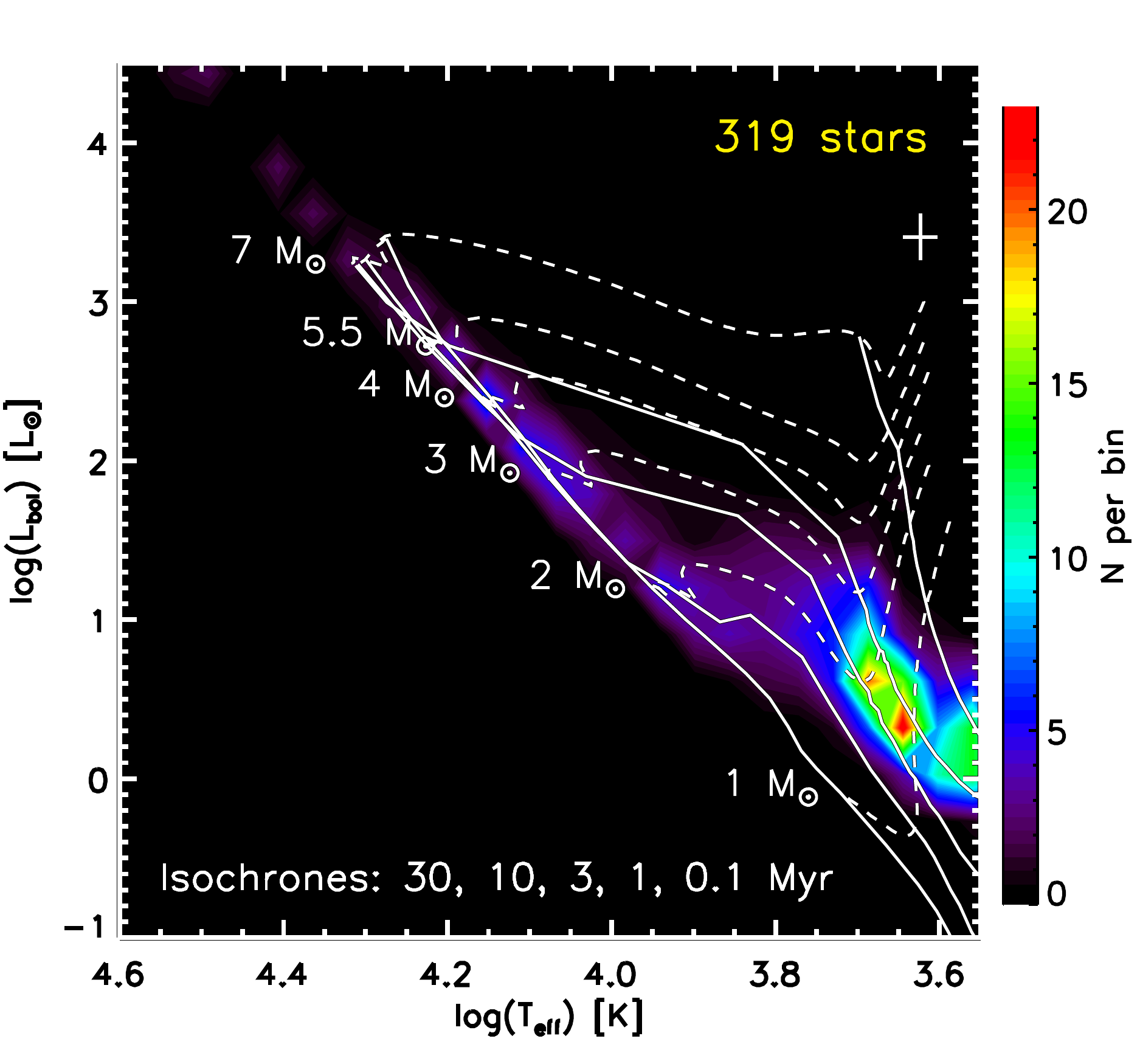}
\includegraphics[width=0.8\columnwidth]{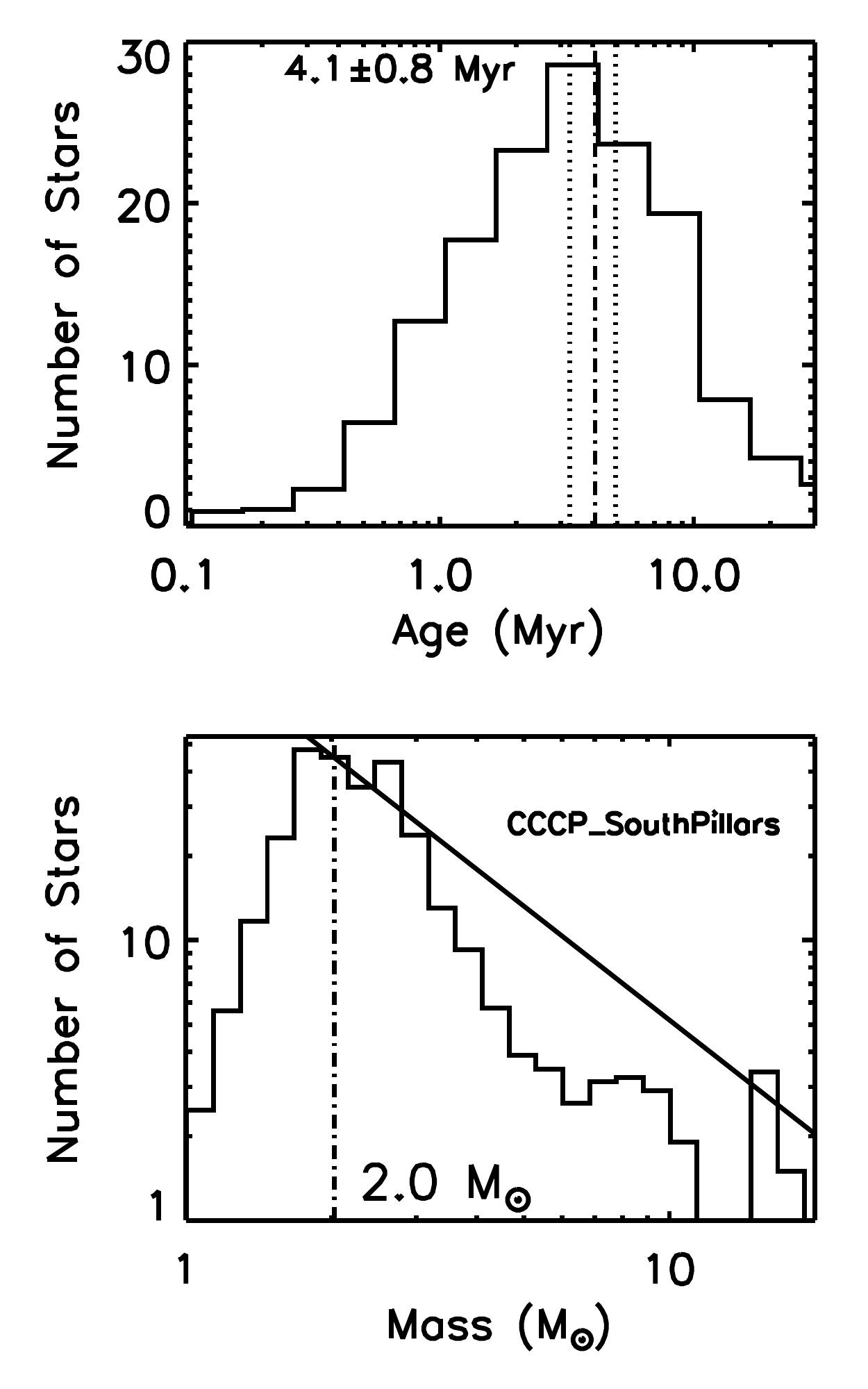}
\caption{{\em Top to bottom:} pHRD, $t_{\star}$ and $M_{\star}$ distributions for CCCP Region C, the South Pillars.
  \label{fig:RegC}
}
\end{figure}
Most of the ongoing star formation traced by {\em Spitzer} YSOs in the CNC occurs in the South Pillars region, where feedback from massive stars continues to sculpt the remaining giant molecular clouds \citep{SP10, CCCPYSOs, CarinaHerschel}. The X-ray bright, diskless population, however, reveals that the evident embedded population intermingles with a significantly more evolved population ($\tau_{\rm SF}=4.1\pm 0.8$~Myr; Figure~\ref{fig:RegC}) that predates the Tr 16 and 14 clusters. Ages varying from 1.1 to 4.3~Myr among the various subclusters in the South Pillars were reported by \citet{AgeJX}. Individual subclusters display varying disk fractions as well, indicated qualitatively by the varying counts of YSOs versus diskless sources in each one (red versus goldenrod dots in Figure~\ref{fig:overview}). Bochum 11 has few associated YSOs, Collinder 228 has relatively more, and an unnamed, obscured cluster to the immediate Southwest of the Treasure Chest hosts many YSOs. The Treasure Chest itself is perhaps the youngest embedded cluster in the CNC \citep{Smith+TC05} but its density and high MIR nebulosity precluded detection of its YSO population using {\em Spitzer}.

\begin{figure}[thb]
\centering
\includegraphics[width=\columnwidth]{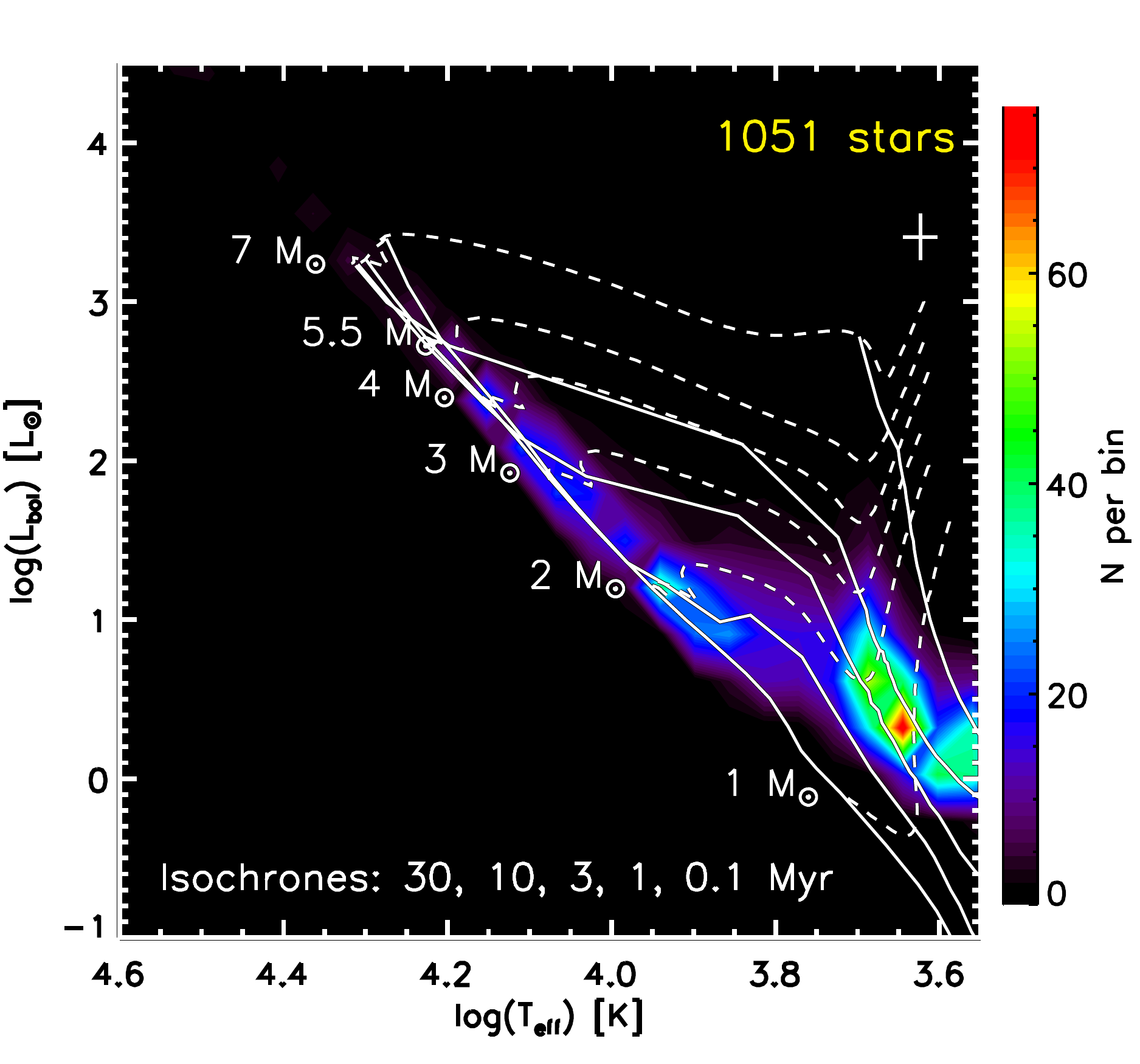}
\includegraphics[width=0.8\columnwidth]{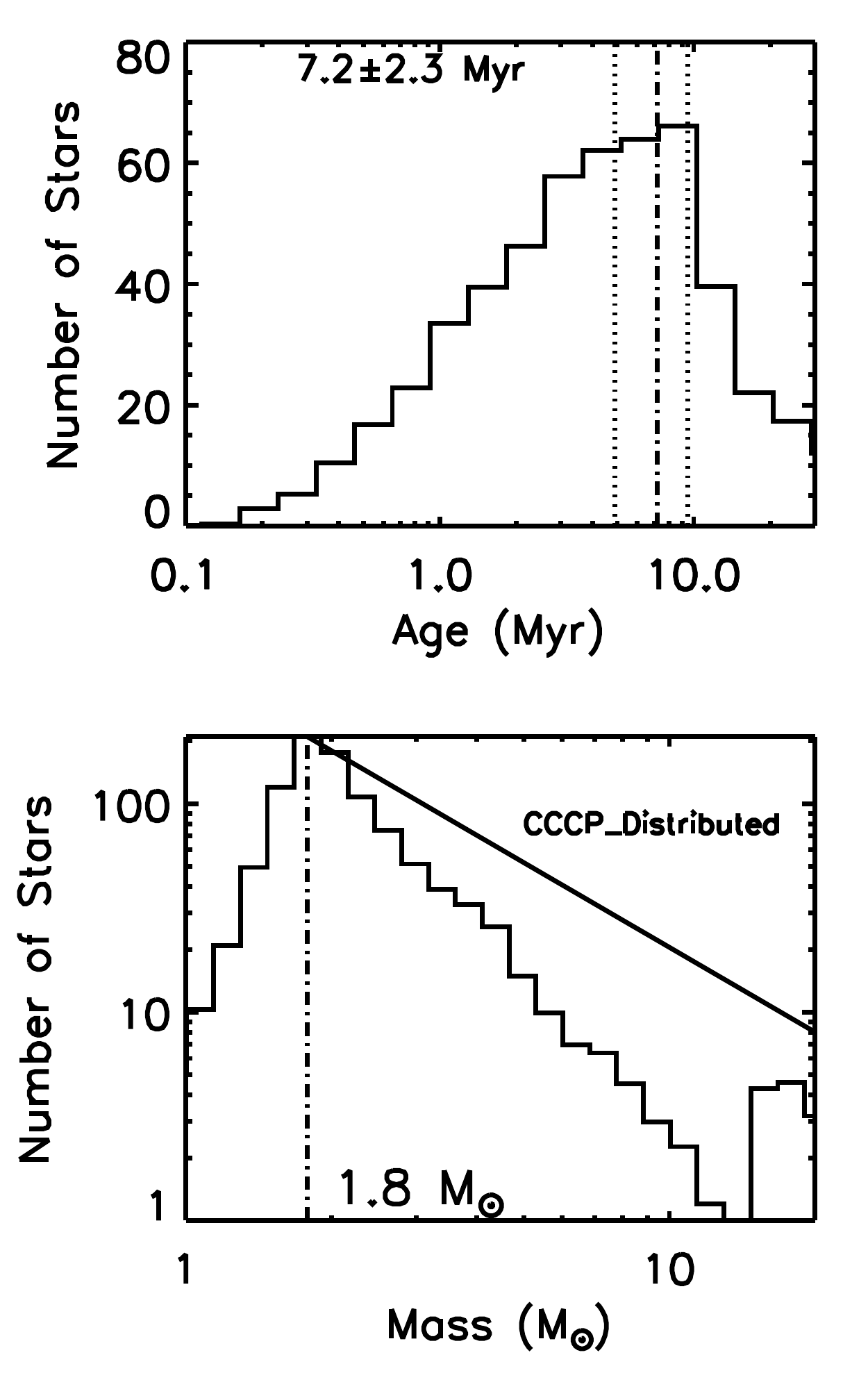}\\
\caption{{\em Top to bottom:} pHRD, $t_{\star}$ and $M_{\star}$ distributions for CCCP Region D, the distributed population.
  \label{fig:RegD}
}
\end{figure}
%%%%%%%%%%%%%%%
Half of the CCCP stellar members lie outside of the three main spatial overdensities in a distributed population \citep[Region D;][]{CCCP_xclust}. 
Not all of this population is truly distributed; some localized overdensities have been grouped into small subclusters \citep{K14_clust,Buckner+19}. 
YSOs mixed within the distributed population and its most obvious smaller subclusters (Figure~\ref{fig:overview}) reveal that region D traces a wide range of ages, likely spanning the full star-forming history of the CNC.
This age range is reflected in the relatively broad peak of the $t_{\star}$ distribution in Figure~\ref{fig:RegD}. Compared to the other sub-regions, we have decreased the histogram binsize and adjusted the break point, yielding $\tau_{\rm SF}= 7.2\pm 2.3$~Myr for Region D.
We find that the duration of start formation in the CNC distributed population, while dominated by the most evolved stars in our CCCP IR-bright, diskless sample, is comparable to the age of the most evolved massive cluster, Tr 15.

%%%%%%%%%% TABLE 2 %%%%%%%%%%%%%
 \begin{table*}
% \tabletypesize{\small}
% \tablewidth{0 pt}
% \tablecaption{ \label{tab:results}
 \caption{ \label{tab:results}
   Duration of Star Formation in the Various CCCP Sub-Populations}
\begin{center}
 \begin{tabular}{lccccccc}
   %\tablehead{
   Region/sample & $N$ & $N_{t\star}$ & $\tau_{\rm SF}$~(Myr)
    & $M_{C}$~(\Msun) & $N_{C}$ & $N_{\rm IM}$ & $f_{\rm XIM}$ \\ % }
    % \startdata
   \hline\hline
 Gaia-ESO ($T_{\rm eff,S}$) & \edit1{139}\tablenotemark{a} & \edit1{99} & $2.7\pm0.8$ & 2.3 & \edit1{79}  & \edit1{152} & \edit1{0.78} \\
 Gaia-ESO                & \edit1{139} & \edit1{88} & $2.7\pm0.8$ & \edit1{2.7} & \edit1{88} & \edit1{145} & \edit1{0.94}  \\
 \hline
 Tr 15                   & 253 & 124 & $6.5\pm1.3$ & 1.8 & 203 & 335 & 0.60  \\
 Tr 14/Cr 232                   & 356 & 194 & $2.6\pm0.5$ & 3.0 & 124 & 353 & 0.85  \\
 Tr 16                   & 290\tablenotemark{b} & 144 & $2.6\pm0.5$ & 3.0 &  85\tablenotemark{b} & 280\tablenotemark{b} & 0.80  \\
 South Pillars           & 319 & 156 & $4.1\pm0.8$ & 2.0 & 191 & 297 & 0.81  \\
 Distributed             & 1050 & 550 & $7.2\pm2.3$ & 1.8 & 760 & 1184 & 0.64 \\
 \hline
 All                     & 2269 & 1168 & $3.6\pm0.6$ & 1.8 & 1651 & 2118 & 0.77 \\
\hline
 % \enddata
 \end{tabular}
 \end{center}
 \tablecomments{$N$ is total number of stars in region/sample,\deleted{ $M_{C1}$} and \edit1{$N_{t\star}$ gives the number of model stars included in the age distribution used to determine $\tau_{\rm SF}$. $M_{C}$ is the mass} cutoff \deleted{for the first and }\edit1{where} the second iteration of the modeled mass functions \edit1{turns over from a Salpeter slope},\deleted{ while $N_{C1}$} and $N_{C}$ is the equivalent number of stars more massive than this turnover mass. $N_{\rm IM}$ gives the number of stars with $M_{\star}\ge 1.8$~\Msun\  predicted by scaling a Kroupa IMF to the final mass function, and $f_{\rm XIM}$ is the fraction of $N_{\rm IM}$ stars that were detected in X-rays.}
 \tablenotetext{a}{This sample includes 14 early-type stars with no $T_{\rm eff,S}$ reported by D17, which we modeled using only our $W_i(\chi^2)$ weighting function.}
 \tablenotetext{b}{The Tr 16 region sample size and inferred stellar population has been significantly reduced by the photometric zone-of-avoidance around $\eta$~Car.}
 \end{table*}

\section{Discussion}\label{sec:discussion}

\subsection{The Complicated Star Formation History of the Carina Nebula Complex}\label{sec:itscomplicated}

Our $\tau_{\rm SF}$ results (Table~\ref{tab:results}) reveal that star formation began at different times at different locations within the CNC, producing a
stellar population
with a hierarchical, multi-clustered structure.
Various locations within the CNC continue to host ongoing star formation, as evidenced by protostellar populations \citep{CCCPYSOs,CarinaHerschel}. Our age distributions (with the possible exception of Region A1) would be populated to the lowest measurable $t_{\star}$ if YSOs with IR excess emission were included.

\begin{figure}[thb]
\centering
\includegraphics[width=\columnwidth]{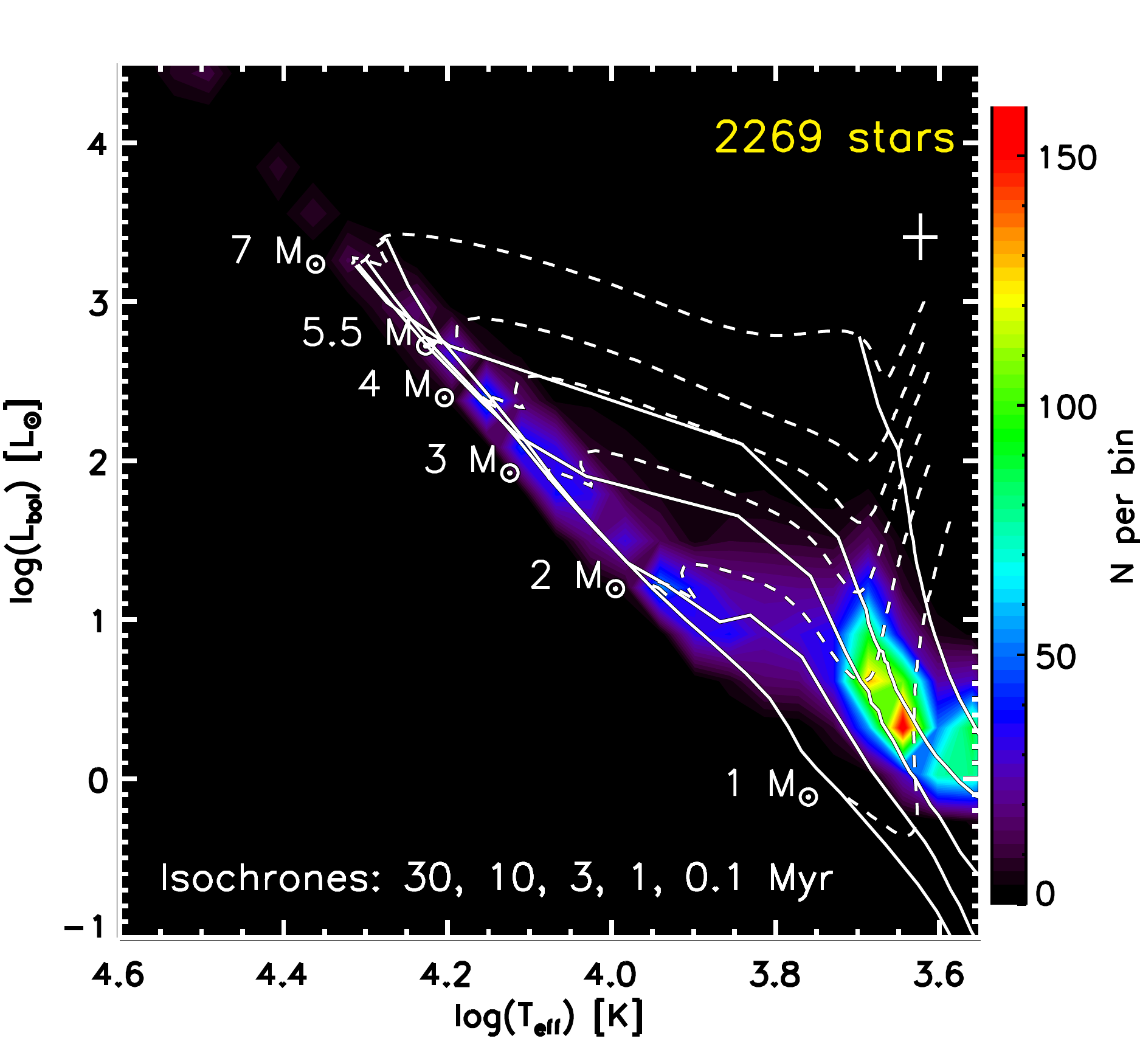}
\includegraphics[width=0.8\columnwidth]{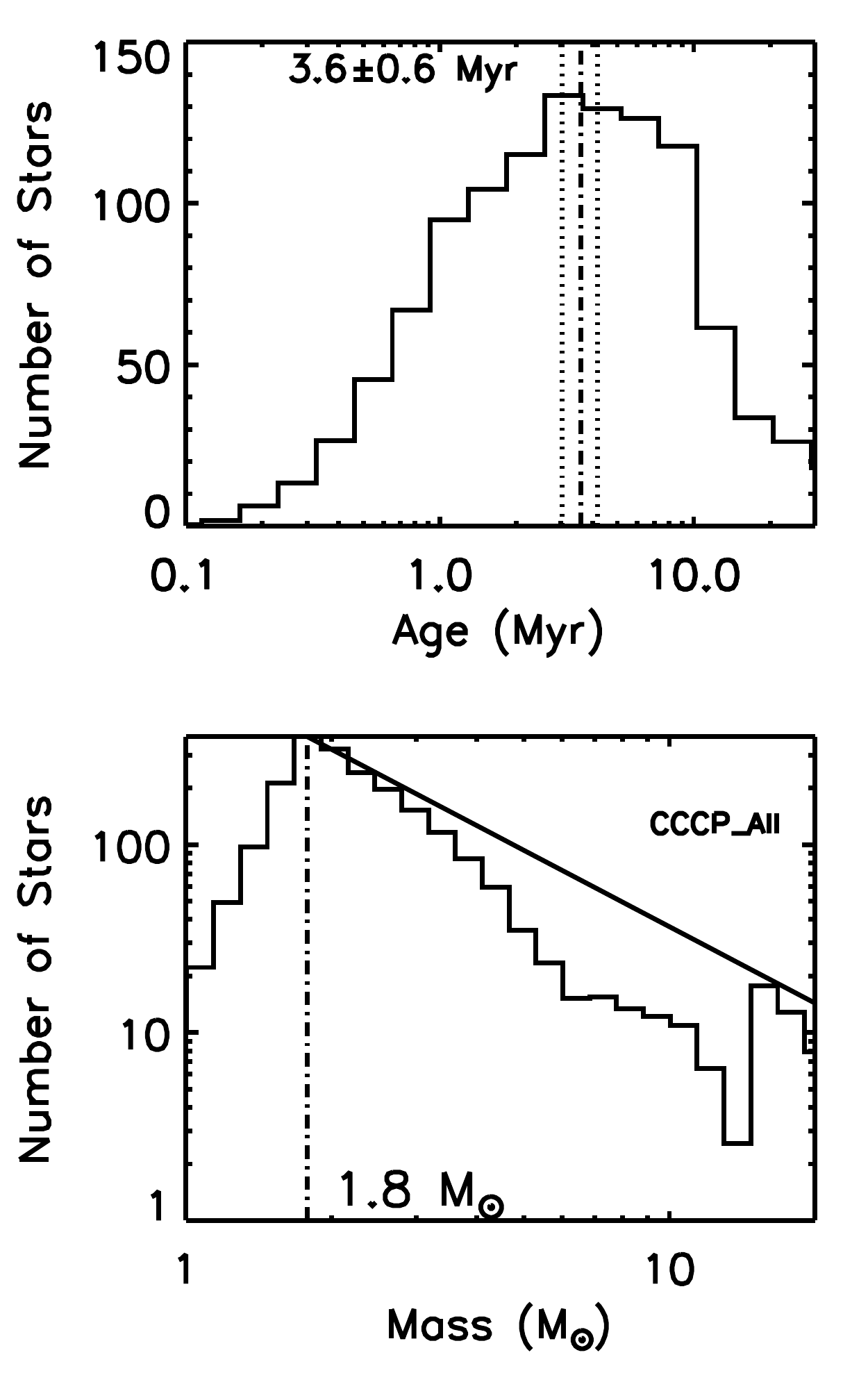}
\caption{{\em Top to bottom:} pHRD, $t_{\star}$ and $M_{\star}$ distributions for all CCCP X-ray selected, diskless members with valid SED fits. To produce the mass function shown here, no single isochronal age value was appropriate, so we instead aggregated the $M_{\star}$ distributions from each constituent sub-region (Figures~\ref{fig:RegA1}--\ref{fig:RegD}).
  \label{fig:AllH2}
}
\end{figure}

We can piece together a global star formation history of the CNC (Figure~\ref{fig:AllH2}). The first stars formed 7--10~Myr ago as a GMC with initial mass likely exceeding $10^6$~\Msun\  \citep{G+88,P+12}. %[CITATIONS FOR MOLECULAR MASS].
Extrapolating from the remnant morphology apparent in IR dust maps, the structure of the original GMC was probably dominated by one large filament running from Southeast to Northwest with one or more intersecting, smaller filaments. The initial collapse phase produced, within a few Myr, Tr 15, multiple smaller clusters (including Bochum 11), and a distributed stellar population formed from lower-density regions that were never gravitationally bound. 

The most spectacular collapse phase commenced ${<}3$~Myr ago, as dense gas channeled along the filaments to their main intersection nodes in the center of the nebula rapidly collapsed to form the massive Tr 16\footnote{Tr 16 is in reality a collection of smaller subclusters \citep{CCCP_xclust,K14_clust,Buckner+19} that are approximately coeval \citep{AgeJX}.} and 14 clusters (as well as Collinder 232\footnote{Previous studies at visual wavelengths \citep{Tapia+03,Hur+12} questioned the existence of the Collinder 232 cluster, probably because parts of it are heavily obscured by a dust pillar. Collinder 232 presents distinct overdensities in both X-ray sources and intermediate-mass YSOs (see Figure~\ref{fig:zoom}).}) in rapid succession. Feedback from the radiation and stellar winds of the very massive stars formed in this event (including $\eta$~Car in Tr 16 and the O3.5--4 V((f+)) stars in Tr 14; \citealt{W+02_O2,SB07}), and possibly the onset of supernovae \citep{CCCP_Tr15,CCCP,CCCPDiffuse} from the most massive stars formed in the initial collapse phase sculpted the remaining molecular material, producing the global, bipolar superbubble structure of the Carina Nebula. The remaining molecular filaments were eroded, forming complex structures in the South Pillars and several pillars in the northern regions of the nebula \citep{H+15}. The remnants of the original filament-node structure are apparent in the large, V-shaped dust lane immediately South of Tr 16 and 14, and the Great Pillar further south that points almost precisely toward the vertex of the V \citep{SB08_Handbook}.

The older, distributed population is consistent with dynamical evolution of many smaller sub-clusters over several Myr timescales, perhaps accelerated by the evaporation of dense gas by massive star feedback.
The current, and likely final, phase of star formation in the CNC is underway within the clumpy molecular remnants, possibly triggered by massive star feedback and revealed by very young clusters of YSOs revealed by the retreat of the remaining pillars \citep{SP10}. In the near future, multiple supernova explosions from Tr 16 and 14 may remove the remaining dense gas, destroying the Carina Nebula and halting star formation for good.

Previous studies of the star formation history in the CNC using observations at visual wavelengths \citep[e.g.,][]{DG-E01} were necessarily restricted to just the central, lightly-obscured region containing Tr 14, 15, and 16. But one recent study leveraged CCCP X-ray and NIR datasets to measure stellar ages across a much wider field, probing the more reddened populations. Introducing a new technique called AgeJX, \citet{AgeJX} used a combination of CCCP X-ray data and deep VLT/HAWK-I $JHK_S$ photometry \citep{CCCP_HAWK-I} to derive median isochronal ages for each of 19 small subclusters identified by \citet{K14_clust} plus the unclustered population. The \citet{AgeJX} study of the CNC was therefore restricted to the smaller HAWK-I field that covered the central 25\% of the 1.4~deg$^2$ CCCP survey area. AgeJX is complementary to our methodology in many respects. It uses the same \chandra/ACIS X-ray point-source data, assuming an X-ray spectral shape characteristic of low-mass, T Tauri stars to assign a mass to each star using an empirical $L_X$--$M_{\star}$ relation \citep{P05}. An extinction-corrected $M_J$ is then used to place stars on the same \citet{SDF00} pre-MS isochrones used in our naked SED models. AgeJX was restricted to ages ${<}5$~Myr and low-mass ($M_{\star}<1.2$~\Msun) stars, requiring deep $J$-band counterpart photometry for faint X-ray sources that may additionally lie behind large absorbing columns. The overlap between individual AgeJX sources and our CCCP IR-bright diskless sample is minimal.

The number of stars per subcluster used in the AgeJX analysis of the CNC ranged from 6 to 111, with typical clusters containing 20--40 stars suitable for AgeJX dating. Most of the \citet{K14_clust} subclusters are too small to contain a statistically robust sample of intermediate-mass stars, but each belongs to one of the larger-scale spatial overdensities \citep{CCCP_xclust} analyzed here. Specifically, region A1 (Tr 15 and environs) contains subclusters G, F, H, and I (AgeJX 2.8--4.8~Myr); Region A2 (Tr 14 and Collinder 232) contains subclusters A--D (AgeJX 1.5--2.8~Myr), Region B (Tr 16) contains subclusters E, J, K, and L (AgeJX 2.4--3.6~Myr), and Region C (South Pillars) contains subclusters M, O, P, Q, R, S, and T (AgeJX 1.1--4.3~Myr). The upper bounds of these AgeJX ranges agree very well with our $\tau_{\rm SF}$ for the Tr~14 and South Pillars regions (Table~\ref{tab:results}). The oldest AgeJX measurement in Tr~16 is $3.6\pm0.7$~Myr for subcluster K, which is marginally greater than $\tau_{\rm SF}$. Subcluster K contains $\eta$~Car and is therefore absent from our analysis due to MIR saturation, but we find good agreement between AgeJX and $\tau_{\rm SF}$ for the remaining subclusters in Tr~16.
The truncation of AgeJX at 5~Myr will cause this method to underestimate the true ages of Tr~15 and the distributed population. In the latter case, AgeJX returns an age of 4.0~Myr based on 354 stars, compared to $\tau_{\rm SF}=7.2\pm 2.3$~Myr from 1050 stars (Table~\ref{tab:results}). Part of this discrepancy may be due to the restriction of AgeJX to the central regions of the CCCP survey area, where even stars found to be unclustered may be younger on average than those in the farther reaches of the distributed population.

We conclude that our results are broadly consistent with past estimates of stellar ages in the CNC. Nearly all of the discrepancies previously reported in the literature can be attributed to difficulties in defining membership of individual clusters, a task made even more challenging by the ubiquitous presence of the large and systematically older, distributed young stellar population.

%\clearpage
\subsection{Impact of the Choice of Pre-MS Evolutionary Models}\label{sec:evochoice}

A number of systematics, in particular neglect of stellar accretion histories and treating binary systems as single stars, both of which we have done here, are known to contribute to apparent isochronal age spreads of ${\sim}10$~Myr on HRDs of very young (${<}20$~Myr) star-forming regions (see \citealp{S14} for a thorough review). That said,
we can be reasonably confident in {\it relative} measurements of isochronal ages among different stellar subpopulations, and the observed intrinsic age spread across the global CNC stellar population, with its numerous subclusters, is real.
The accuracy of {\it absolute} isochronal ages, however, is a matter of ongoing investigation and debate.
%, as different pre-MS models can predict discrepant ages for the same stars by factors of ${\sim}2$ [CITATION?].
In many evolutionary models, including those used in this study, a single isochrone fails to connect intermediate-mass stars to subsolar-mass pre-MS stars in the same population, predicting younger ages for lower-mass stars \citep[see, e.g.,][]{HW04,DH+19}.

Relatively few pre-MS evolutionary models provide detailed calculations for the ${>}2$~\Msun\ stars that dominate our CCCP IR-bright sample (most exclude this mass range altogether to focus on low-mass pre-MS evolution). \citet{DH+19} compare the results of placing the three most massive stellar components of eclipsing binary systems (1.4, 2.6, and 5.6~\Msun, respectively) in the Upper Scorpius OB Association on isochrones from the Dartmouth \citep{Dartmouth}, MIST \citep{Choi+16_MIST,Dotter16_MIST}, and PARSEC \citep[v1.0;][]{PARSEC_v1.0} evolutionary models. These authors reported good agreement among all isochrones for the ages of these intermedate-mass stars, and while the older \citet{SDF00} isochrones were not considered, the shapes and locations of their pre-MS isochrones are very similar to the three newer model sets analyzed by \citet{DH+19} across the 2--6~\Msun\ mass range.

The pre-MS isochrones and evolutionary \edit1{tracks} adopted in our naked SED models \citep{BM96,SDF00}, in spite of their advanced age, still give reasonable results for intermediate-mass stars. The most promising new grid of evolutionary models appears to be the recent extension of the Geneva evolutionary tracks to the pre-MS phase by \citet{H+19}. These models include a detailed treatment of accretion history and computation of the stellar birthline across the entire range of present-day stellar masses. They predict that stars of ${<}3$~\Msun\ are fully-convective when accretion ends, which is precisely what we have observed with our large sample of X-ray bright, diskless 2--3~\Msun\ IMPS in the CNC. The locations and shapes of the evolutionary tracks and isochrones for $t_{\star}\ge 1$~Myr as well as the ZAMS arrival times reported by \citet{H+19} for 2--5~\Msun\ stars are very similar those implemented in our models.

Some recent models for {\it low-mass} pre-MS evolution introduce the effects of magnetic fields (magnetic pressure support or large starspot covering fractions) that inhibit the contraction of fully-convective stars on Hayashi tracks, increasing the isochronal ages of ${<}1$~\Msun\ pre-MS stars by factors of ${\sim}2$ \citep{SP15,Dartmouth15,Feiden16,Jeffries+17}. Intrinsic, magneto-coronal X-ray emission characteristic of the cool, convective IMPS in our sample (Nu\~{n}ez et al. in preparation) raises the question of whether a similar correction for isochronal ages of IMPS might be warranted. We suspect any such correction would be small, however, because a full factor of 2 increase in our reported ages would place the birth of the massive Tr 14 and 16 stellar clusters at 5--6~Myr ago, which exceeds the lifetimes of their most massive stars. This would, in turn, require the massive stars in Tr 16 and Tr 14 to form ${\ga}2$~Myr later than the intermediate-mass stars in the same (sub)clusters.
%It would also potentially increase the core-halo age gradient observed in Tr~14 \citep{A+07_Tr14,AgeJX}. 

Uncertainties in pre-MS isochronal ages provide the most important systematic uncertainty in our results, but based on our review of the current literature this systematic appears to be considerably smaller than the factors of ${\sim}2$ differences between the magnetic and non-magnetic low-mass isochrones.

\subsection{IMPS and the ``X-ray Desert''}\label{sec:desert}
Nu\~{n}ez et al.\ (2019, in preparation) present a spectral fitting analysis of the several hundred brightest CCCP X-ray sources (excluding OB stars) and found that those with IR counterparts classified as IMPS on the pHRD were systematically more luminous in X-rays compared to lower-mass T Tauri stars or intermediate-mass AB stars. IMPS extend the convective T Tauri star relation of $L_X/L_{\rm bol}\sim 10^{-3.6}$ \citep{P05} to higher X-ray luminosities ($L_X\sim 10^{32}$~erg~s$^{-1}$). The strongest stellar coronal X-ray flares may thus be produced by IMPS.
However, after IMPS complete their descent of the convective Hayashi tracks, they develop radiative cores that grow rapidly, sending stars horizontally across Henyey tracks toward the ZAMS (producing the above-mentioned R-C gap described by \citealp{M+07}, which is apparent on all of our pHRDs but most pronounced for Tr~14 and 16; see Figures~\ref{fig:RegA2} and \ref{fig:RegB}, respectively).
% This transition is relatively quick, opening a gap on the HRD (analogous to the well-known Hertzsprung gap observed between the MS and the red giant branch in globular clusters, \citealp{M+07,M10}).
The growth of the radiative core drastically alters the magnetic field structure of the star \citep{G+12}.
%The powerful, intrinsic X-ray emission from IMPS is consistent with a scaled-up version of T Tauri convective dynamos driving magneto-coronal flares, but only at early ages.
Unlike the case for low-mass stars, the IMPS dynamo-driven X-ray emission decays rapidly after the R-C transition \citep{G16}, and should disappear completely prior to arrival on the fully-radiative ZAMS.

Because a greater fraction of intermediate-mass stars reach the ZAMS and become X-ray dark as a population ages, we expect that the magnitude of this deficit,  (the ``aridity'' of the X-ray desert) will increase as a function of $\tau_{\rm SF}$.\footnote{For the youngest stellar populations we expect a mass distribution consistent with Salpeter for X-ray selected, diskless IMPS, as observed by P16 for the IR dark cloud M17 SWex.}
To explore this effect, for the mass distribution of each region/sample in Table~\ref{tab:results} we define a parameter $f_{\rm XIM}$, the fraction of X-ray detected intermediate-mass stars with $M_{\star}\ge 1.8$~\Msun.
We estimate the expected number $N_{\rm IM}$ of intermediate-mass stars in the parent population  by integrating the Salpeter IMF (scaled to the value of the $M_{\star}$ distribution at the cutoff mass $M_{C}$).
After correcting for the varying mass-incompleteness in each $M_{\star}$ distribution in the range 1.8~\Msun~$\le M_{\star}< M_{C}$ using the same scaled Salpeter IMF, we integrate over this corrected distribution to find $f_{\rm XIM}$.

%\clearpage
\begin{figure}[thp]
%%\epsscale{1.}
%\centering
\includegraphics[width=\columnwidth]{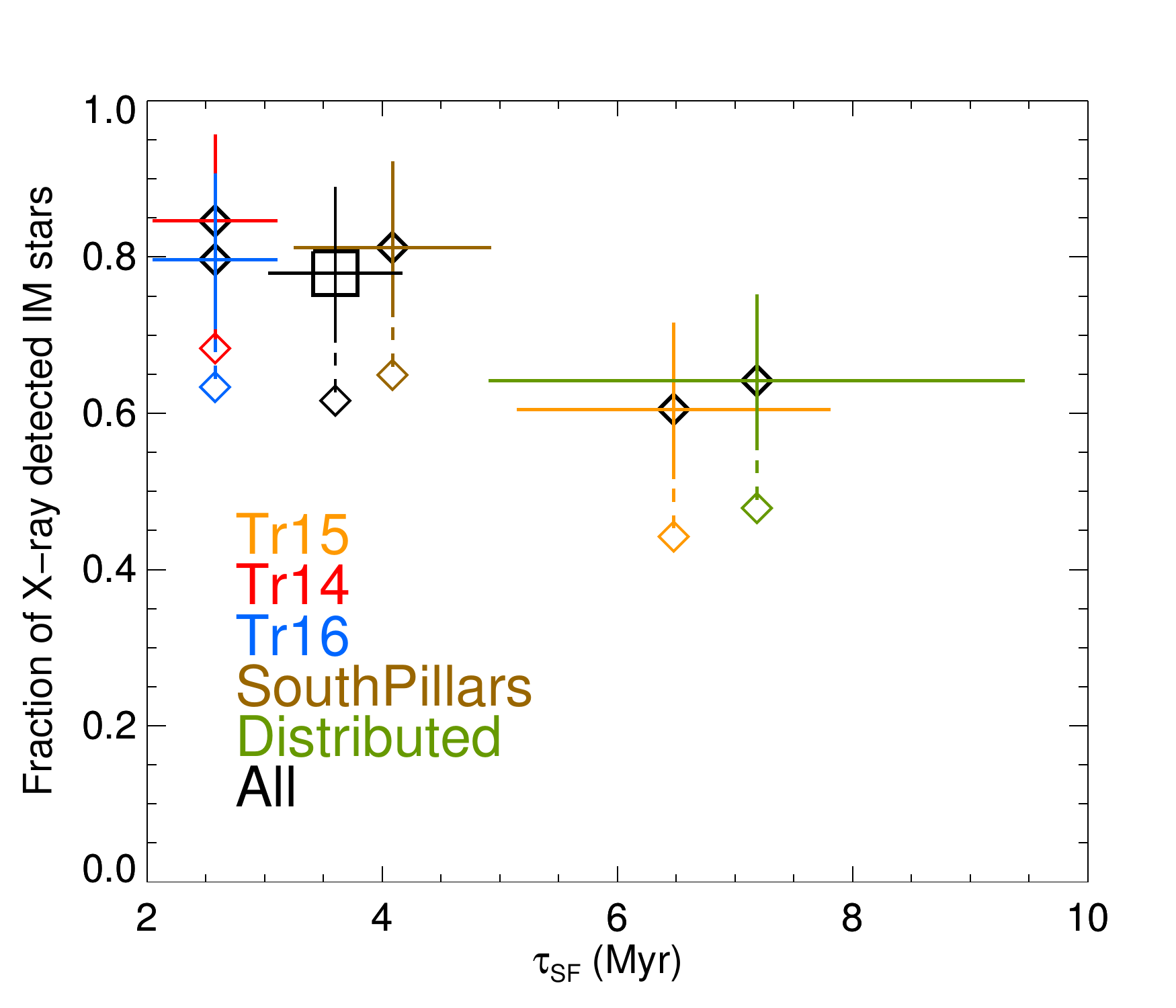}
\caption{Trend of decreasing $f_{\rm XIM}$ with increasing $\tau_{\rm SF}$. The   dashed lines and lower,
  colored diamonds show the correction for the systematic offset
  introduced by the intrinsic spread in the modeled mass distributions.
  \label{fig:xfrac}
}
\end{figure}
As expected, we observe a trend of decreasing $f_{\rm XIM}$ with increasing $\tau_{\rm SF}$ (Figure~\ref{fig:xfrac}; error bars on $f_{\rm XIM}$ are estimated assuming Poisson counting statistics for each bin in the $M_{\star}$ distribution).
It would be tempting to use this qualitative trend to infer the real binary fraction among intermediate-mass stars, for example, but systematic effects preclude us from drawing such firm, quantitative conclusions. One complication is that $N_{\rm IM}$ provides only a lower limit on the true stellar population, because many intermediate-mass stars in the CNC still have disks and were hence not counted among our sample. Furthermore, we have no information about the frequency at which real binary companions are detected as X-ray point sources, the sensitivity of which varies across the large CCCP mosaic \citep{CCCPCat}. But the systematic most relevant to the current analysis involves the shape of the $M_{\star}$ distributions themselves, given the challenges of precisely constraining stellar masses from SED modeling. For the Gaia-ESO spectroscopic comparison sample, we find a deeper deficit in intermediate-mass stars with the more stringent $P(T_{\rm eff,S})$ constraints on the SED models than with our standard $P(\tau_d,\tau_x)$ weighting ($f_{\rm XIM}=\edit1{0.78}$ versus \edit1{0.94}; see Figure~\ref{fig:pMass_D17} and Table~\ref{tab:results}). This is purely an artifact of limited precision in measuring stellar masses via broadband photometry alone. A tentative correction for this effect, simply  $f_{\rm XIM} - \edit1{0.16}$,
is also illustrated in Figure~\ref{fig:xfrac}. We caution that this correction is not strictly appropriate, given the additional selection criteria imposed by D17 and by us on the Gaia-ESO comparison sample that were not imposed on the full CCCP IR-bright sample, but it does extend the uncertainty on $f_{\rm XIM}$ to encompass a more plausible range of values.

%We hope that the qualitative trend of an increasingly arid X-ray desert at intermediate-masses with increasing age provides a useful illustration and warning about hidden biases that can affect X-ray selected stellar samples.

\section{Conclusions}\label{sec:conclusions}

We have presented a pilot study in which we have developed a novel methodology for constraining the duration of star formation and applied it to multiple young stellar populations observed in the Carina Nebula Complex (CNC).
We employ the results of IR SED fitting to individual X-ray-selected, diskless intermedate-mass stars to synthesize models of the ensemble population and probe distributions of stellar masses and ages. Because IMPS rapidly evolve along radiative tracks to become late B- and A-type stars on the ZAMS (including Herbig Ae/Be stars), they serve as the most sensitive available chronometers for the first ${\sim}10$~Myr in the evolution of massive stellar clusters and associations.

Our SED modeling methodology uses all available photometric data simultaneously, incorporating external age constraints based on the lack of mid-IR excess emission and the presence of X-ray emission that are communicated to the model via weighting functions. We treat each star as a cloud of probability in SED model parameter space, analyzing the $\log{L_{\rm bol}}$--$\log{T_{\rm eff}}$ plane (pHRD) coupled with mass and age distributions.  This avoids the pitfalls of interpreting individual photometry data points with gaussian error bars, which can be large and are inherently asymmetric with respect to non-linear isochrones and evolutionary tracks. We validate our technique using a comparison sample of \edit1{139} stars with spectroscopic $T_{\rm eff,S}$ measurements or broad constraints from the Gaia-ESO survey (D17), and find that we can successfully identify the correct isochronal age from the weighted SED models fit to broadband IR photometry alone. This validation exercise provides a road map for what we would consider a ``gold standard'' for observational studies of young stellar populations in massive Galactic star-forming regions suffering from severe differential extinction: visual/NIR spectroscopy to measure $T_{\rm eff,S}$ for individual stellar sources combined with IR SED modeling analysis to tightly constrain $L_{\rm bol}$ and reddening. 

Our study encompasses the entire 1.4~deg$^{2}$ field of the CCCP, including the substantial fraction of distributed and/or significantly reddened stars in the CNC, providing a homogeneous analysis of wide-field X-ray and IR imaging data.
We find that star formation commenced throughout the CNC ${\sim}10$~Myr in the past \citep[as proposed by][]{DG-E01}, after which the SFR rapidly accelerated to produce the massive Tr 15 cluster ${\sim}6.5$~Myr ago and peaked 2--3~Myr ago with the birth of the very massive Tr 16 and 14 clusters, which contain some of the most massive stars known in the Galaxy. Our results generally agree with isochronal ages reported previously for various constituent CNC (sub)clusters \citep[e.g.,][]{FFM80_Tr15,Carraro02,Tapia+03,A+07_Tr14,Hur+12,AgeJX}. 

Our X-ray selection criterion imposes unique selection biases on the resultant stellar samples. In particular, the powerful coronal X-ray emission produced by convective IMPS disappears once these objects reach the ZAMS along radiative tracks \citep[][Nu\~{n}ez et al.\ in preparation]{G16}. We observe the effects of this selection bias as a deficit of intermediate-mass stars in our modeled mass distributions with respect to a standard Salpeter IMF slope, the magnitude of which increases with age of the stellar population. Late B- and A-type stars (including Herbig Ae/Be stars) near the ZAMS can also be included in our samples if they have lower-mass, convective stellar companions producing detectable X-ray emission.

Subsequent papers in this series will implement our methodology for a uniform, comparative study of the duration of star-formation in more than two dozen other Galactic massive star-forming regions for which comparable X-ray and IR photometric data exist. The results of this extended study will enable improved measurements of star-formation rates, star-forming efficiencies, and the dynamical evolution of giant \hii regions.

\acknowledgements
We thank M. A. Kuhn and L. A. Hillenbrand for numerous discussions and suggestions that substantially improved this work. \edit1{We are grateful to the anonymous referee for  helpful comments and suggestions that improved the interpretation of the spectroscopic comparison sample and the explanation of our methodology.} This research was supported by the NSF through grant CAREER-1454333 (PI M. S. Povich). JTM and EHN acknowledge support from the Cal-Bridge program through NSF awards DUE-1356133 and AST-1559559. The scientific results are based in part on observations made by the {\em Chandra X-ray Observatory} and published previously in cited articles. This work is based in part on archival data obtained with the {\em Spitzer Space Telescope}, which is operated by the Jet Propulsion Laboratory, California Institute of Technology under a contract with NASA. 
This publication makes use of data products from the Two Micron
All-Sky Survey, which is a joint project of the University of
Massachusetts and the Infrared Processing and Analysis
Center/California Institute of Technology, funded by NASA and the NSF. 

%{\em Facilities:}
\facility{CXO (ACIS)}, \facility{Spitzer (IRAC)},  \facility{CTIO:2MASS}

%%%%%%%%%%%%%%%

\end{document}